\documentclass[mnsc,nonblindrev]{informs3h}   

\usepackage{natbib, graphicx, color, tabularx, mathtools, bbold, soul, bbm, enumerate, eurosym, multirow, booktabs, xcolor,subfig}
\usepackage[normalem]{ulem}
\bibpunct[, ]{(}{)}{,}{a}{}{,}%
\def\bibfont{\small}%
\usepackage{epstopdf}

\newtheorem{appxproposition}{Proposition}[section]

\TheoremsNumberedThrough     

\EquationsNumberedThrough    

\begin{document}

\RUNTITLE{Generative AI and Copyright}

\TITLE{Generative AI and Copyright: A Dynamic Perspective}


\RUNAUTHOR{Yang and Zhang}
\ARTICLEAUTHORS{%
\AUTHOR{S. Alex Yang}
\AFF{London Business School,  \EMAIL{sayang@london.edu}}
\AUTHOR{Angela Huyue Zhang}
\AFF{University of Southern California Gould School of Law, \EMAIL{azhang@law.usc.edu}
}
} 

\ABSTRACT{The rapid advancement of generative AI is reshaping the creative industry. Yet its future development and applications hinge crucially upon two copyright issues: (1) whether and to what extent creators should be compensated when their content is used to train generative AI models (the fair use standard), and (2) whether AI-generated content should qualify for copyright protection (AI-copyrightability). The paper investigates the economic implications of these two regulatory issues and their interactions. We develop a two-period model with endogenous content creation and AI model development to study how fair use standard and AI-copyrightability jointly influence AI development and welfare for different stakeholders. Our analysis reveal novel inter-temporal tradeoffs. For example, while generous fair use (use data for AI training without compensating the creator) benefits all parties when abundant training data exists, it can harm social welfare and AI development as such data becomes scarce. We find similar tradeoffs for AI-copyrightability and show that the two policies interact in non-trivial ways: When pre-existing training data are scarce, generous fair use may be preferred only when AI-copyrightability is low. Policy objectives emerge as the primary determinant of optimal AI-copyrightability, whereas the optimal fair use standard is more sensitive to changes in policy goals, technological conditions, and market growth. Our findings underscore the need for policymakers to embrace a dynamic, context-specific approach in making regulatory decisions and provide insights for business leaders navigating the complexities of the global regulatory environment.

}

\KEYWORDS{Generative AI, copyright, fair use standard, intellectual property, creative industry, creator economy, authorship}
\HISTORY{First draft: 4 February 2024; this draft: 5 December 2025.}

\maketitle

\section{Introduction}\label{sect:intro}

The rapid advancement of generative artificial intelligence (AI) is fueling the next wave of productivity. Goldman Sachs has estimated that generative AI could drive a 7\% (or almost \$7 trillion) increase in global GDP and lift productivity growth by 1.5 percentage points over a 10-year period \citep{goldman2023generative}. Among the sectors most likely to be reshaped is the creative industry \citep{de2023generative}. Yet, alongside this surge of optimism, it has also sparked significant legal and economic conflicts, particularly between AI firms and content creators. 

The first major controversy concerns whether the training of AI models qualifies as fair use, which would exempt AI firms from compensating copyright owners for using their content in AI training (``fair use standard"). This issue sits at the heart of numerous intellectual property (IP) infringement litigations filed by publishers, artists and authors against leading AI firms. For example, Getty Images sued Stability.AI for allegedly training its models on Getty’s images without authorization \citep{brittain2023getty}. The New York Times has also filed a high-profile suit against OpenAI and Microsoft, seeking billions in damages for purported copyright violations \citep{grynhaum2023foundation}. In the music sector, major publishers -- including Universal Music Group and Concord -- have sued Anthropic for allegedly using copyrighted lyrics to train its Claude models \citep{brittain2025music}. Meanwhile, the European AI Act requires developers to disclose a summary of copyrighted materials used in model training \citep{eu2024artificial}. Such a transparency mandate could further heighten litigation risks for AI firms. Although the legal outcomes and regulatory frameworks may take years to resolve, the mounting litigation and regulatory pressure is already reshaping industry practice. Leading AI firms are increasingly entering content-licensing agreements to secure training data and mitigate legal exposure. OpenAI, for example, has signed deals with the Associated Press and Axel Springer to license both archived articles and new stories \citep{obrien2023chatgpt,thomas2023axel}, and it has reportedly offered \$1-5 million annually to license news articles for training its AI models \citep{david2023openai}. Google has similarly negotiated multimillion-dollar licensing arrangements with Reddit and various news publishers to support the development of its Gemini models \citep{tong2024reddit}. 

The other major legal controversy concerns whether AI-generated content should qualify for copyright protection (“AI-copyrightability”). In the United States, both the Copyright Office and district courts have repeatedly denied protection to AI-generated work \citep{geiger2023elaborating,brittain2025US}. These decisions have sparked strong backlash, including lawsuits filed by artists challenging the Office’s refusal to register AI-generated creation \citep{brittain2024artist}. This U.S. approach stands in sharp contrast to developments in China. In late 2023, the Beijing Internet Court issued a landmark decision holding that an image created using text prompts in Stable Diffusion could be copyrighted because the user’s prompt design and iterative adjustments reflected original human creative input \citep{wininger2023beijing}. In 2025, the Changshu People's Court reached a similar ruling, granting copyright protection to another AI-generated image and ordering the defendants to pay RMB 10,000 for economic losses and related expenses \citep{chatterton2025changshu}. Emerging empirical evidence underscore the broader economic stakes of these developments. Following a U.S. district court decision declaring AI-generated art uncopyrightable, prices for AI-generated art projects on major freelancing platforms fell by approximately 32\% \citep{li2025ai}.  

Both the issues of fair use and AI-copyrightability pose tremendous challenges to the existing copyright doctrines, igniting vigorous debate among scholars \citep{lemley2023generative, geiger2023elaborating}. Some argue that only human-generated content should be eligible for copyright protection, whereas others believe that AI-created content could be copyrighted if human input plays a central role in its creation \citep{grimmelmann2015there, burk2020thirty, abbott2022disrupting}. However, the extent of human input required remains highly contentious. Likewise, scholars note that the transformative use of copyrighted materials is often considered fair use under US law, but this argument becomes less tenable when AI-generated content is substantially similar to the original copyrighted materials \citep{henderson2023foundation}. 

Beyond these doctrinal complexity, these two issues carry substantial business and policy implications. A primary concern is that restricting the use of copyrighted materials for training AI models will significantly increase data acquisition costs, potentially showing innovation in the AI industry. Conversely, media experts and policymakers warn that failing to adequately compensate content creators may undermine the creative industry \citep{henshall2023experts}. Similar considerations arise in the AI-copyrightability debate. In a public interview, the presiding judge in one of the Chinese cases argued that granting copyright protection to AI-generated content could encourage the use of AI, thereby driving investment and fostering the industry's development.\footnote{\url{http://tinyurl.com/5n84vyx6} (in Chinese)} These competing concerns were reflected in the U.S. Copyright Office’s 2023 Notice of Inquiry on AI and copyright, which elicited more than 10,000 public comments. The Office subsequently released a comprehensive report on copyright and artificial intelligence \citep{USCO2025AI}, underscoring the importance of both copyrightability (Part 2) and the fair-use framework (Part 3) in shaping the copyright ecosystem in the era of generative AI. 

Motivated by the above debates, this paper provides an analytical investigation of how fair use and AI-copyrightability influence the incentives and behaviors of AI companies and content creators, and how these policies affect their income and consumer surplus. Beyond examining their individual effects, we also explore how these two policies interact. Traditionally -- including in the U.S. Copyright Office's recent report --these two issues have been analyzed separately. Yet this separation risks overlook a defining feature of the generative AI content-creation ecosystem. Unlike traditional creative industries, AI companies plays an important role at both sides of the market: on the demand side, they need to acquire data for model training, the cost of which is determined by the fair use standard; on the supply side, the value of the AI service they provide depends on that of the output of these service, which is affected by AI-copyrightability. Recognizing this connection, the paper examines how these policies operate jointly and what their combined economic impacts may be.


To answer the above questions, we develop a two-period model consisting of a regulator, a generative AI company, and a group of content creators. The regulator makes two copyright-related decisions: 1) how {\em generous} the fair use standard is (the amount the AI company need to compensate copyright owners for using their content in model training); 2) the level of AI-copyrightability (the fraction of copyright protection granted to AI-generated content relative to human-created content). Content creators, who are endowed with different skill levels, decide whether to produce content themselves (human content), generate content using a generative AI tool provided by the AI company (AI content), or not produce content at all. Using AI tools lowers the cost of creation, but the quality of AI-generated content is limited by the quality of the underlying model.\footnote{Low-quality models can meaningfully degrade the quality of generated content. For example, CNET paused publication of AI-generated articles after finding that they were ``plagued with errors" \citep{thorbecke2023plagued}.} Content creators earn income by selling content to consumers and (potentially) sell human-generated content to the AI company for model training.\footnote{Motivated by concerns about model collapse \citep{rao2023ai,lutkevich2023model}, we assume that only human content can improve model quality. Our main insights remain qualitatively robust so long as human content is more valuable than AI-generated content for enhancing model performance.} In turn, the AI company earns revenue when creators use its service to generate content and decides how much human-created content to acquire for model training. This decision directly determines the quality of the AI tool available to creators in the second period. To capture the increasingly competitive landscape of the generative AI market, we assume that the AI company jointly chooses the improved model quality and the price of its AI service to maximize the value of its offering to AI creators, subject to a non-negative profit constraint.

Using this model, we show that creators’ content-production decisions follow a double-threshold structure. The most skilled creators continue to produce content themselves; those with intermediate skill levels rely on AI tools to generate content; and the least skilled choose not to produce content at all. These decisions depend on the fair-use standard, the level of AI-copyrightability, and the quality of the AI model. For instance, as model quality improves, the set of creators who choose AI-assisted content creation expands, while the regions corresponding to human-only creation and no creation shrink. This behavioral shift shapes both the demand for the AI model and the amount of human-generated content available for model training. Jointly, they influences the AI company’s incentives to invest in further improving its model.

The interaction between the AI company and content creators depends critically on the amount of data available prior to Period 1 for model training (``pre-existing data”). When pre-existing data are abundant -- typically in domains where model development remains at an early stage (for example, video-generation models) -- the AI company simply acts as a service provider to creators who choose AI-assisted content production. By contrast, when pre-existing training data are scarce -- such as in areas where models are already highly developed and most historical data have been exhausted by prior training (for example, large language models), or where historical data quickly become obsolete -- model improvement relies heavily on newly created content. In this case, the AI company not only provides services to AI-assisted creators, but also serves as a purchaser of human-generated content for model training. As a result, creators' content-production decisions and the AI company's model-development choices are more tightly coupled across periods, creating a richer set of intertemporal strategic interactions.

These relationships have significant implications for the effects of copyright policy. In the setting with abundant pre-existing data, generous fair use (creators are not compensated when their content is used for model training) improves AI model quality, which in turn increases creators’ income and enhances social welfare. Stronger AI-copyrightability always benefits content creators, but its effects on consumers and social welfare are more nuanced. On one hand, it directly harms consumers by enabling creators to capture a larger share of content value; on the other hand, it may indirectly benefit consumers by encouraging more content creation and improving AI model quality. Regarding the interaction between the two, stronger AI-copyrightability can partially substitute the effect of generous fair use in model development. However, the connection between the two policies is relatively weak, and generous fair use remains the welfare-maximizing choice regardless of the level of AI-copyrightability.

By contrast, the effects of these two copyright policies become substantially more complex when pre-existing data are scarce. First, overly generous fair use can impede AI development and reduce social welfare by weakening creators’ incentives to produce human content. Likewise, strong AI-copyrightability directly increases demand for AI tools but simultaneously reduces the supply of human-generated content. In this setting, the interaction between the two policy levers becomes a first-order consideration. For example, generous fair use promotes AI development and enhances social welfare when AI-copyrightability is weak, but it becomes a hindrance to both when AI-copyrightability is strong. 

Finally, we demonstrate that the optimal policy mix is significantly influenced by policy objectives. Interestingly, under the data scarce case, a policymaker prioritizing AI development should set a stricter fair use standard than one that focus on consumer surplus. Further, a policymaker seeking to maximize AI development should adopt a stricter fair use standard than one focused on consumer surplus. Moreover, the optimal fair use standard is more sensitive, relative to AI-copyrightability, to changes in data availability, technological conditions, and market prospects. AI-copyrightability, by contrast, tends to be determined primarily by policy priorities.

As the first analytical study to jointly examine these two major copyright issues in the context of generative AI, our paper offers a framework that captures the nuanced tradeoffs policymakers must balance. In particular, it underscores the need for a dynamic perspective. For example, in determining fair use standard, one must consider not only its short-term effect on model-training costs but also its long-term role in sustaining a supply of human-generated content for future training. Accordingly, we advocate a dynamic, context-specific approach that accounts for regulatory priorities, as well as technological and market conditions. We also caution that policy analysis risks missing critical interactions if fair use and copyrightability are treated in isolation. The connection between these two doctrines highlights the need for regulators -- especially those constrained by existing legal jurisprudence -- to consider their interplay carefully. Finally, our findings suggest that business leaders should anticipate a diverse regulatory landscape and develop strategies to navigate its complexities effectively.


\section{Related Literature}\label{sect:literature}
This paper is related to several strands of literature. First, there is a recent stream of literature on the strategic interaction between different parties in supply chains in the presence of a market of information or data. \cite{bimpikis2019information} highlights an information provider should adjust the quality of information based on the strategic interaction between its customers when selling information to them. \cite{mehta2021sell} examines the data provider's pricing decisions. Also focusing on the setting where a data provider sells data to a buyer who makes operational decisions, \cite{drakopoulos2023providing} establish conditions when it is optimal to provide a free sample to the potential buyer. \cite{gurkan2022contracting} study how a company could better leverage the AI flywheel effect (the virtuous cycle where more usage of an AI model will generate more data that could be used to further improve the model) by delegating certain tasks to an agent subject to moral hazard. \cite{bimpikis2023data} highlight how data generated in one market could be used to affect production market competition in another market by enabling price discrimination. By also focusing on a setting that features data selling (from content creators to AI company) and the consumption of an AI model/data by strategic agents, our research complements the above papers in two dimensions. First, by focusing on generative AI, our model captures a new form of data supply chain, where content creators choose to become the data provider for the AI company or the user of the model/service provided by the AI company. Second, while also capturing the strategic interactions between different players in a data supply chain, our focus is not on identifying optimal decisions for individual companies, but rather on the impact of copyright-related regulations.  

Within this broader theme, our paper is also related to the literature on content aggregation, which examines interaction between content producers (e.g., publishers) and information aggregators (e.g., search engine or social media platforms). \cite{jeon2016news} model the tradeoff between competition (``business stealing") and complementarity (``readership expansion") in this setting, whereas \cite{sismeiro2018competitive, calzada2020news, de2023social} empirically assess the overall impact of information aggregation by accounting for both forces. Although we also study a supply chain in which one party (the AI firm) obtains content from another (content creators), our work differs from this literature in three respects. First, a central tension in content aggregation is competition between aggregators and content creators. By contrast, outputs of generative AI are typically transformative rather than direct substitutes for the underlying works. Consequently, our paper focuses on a different tradeoff: the demand-side incentive that encourages AI firms to improve their models versus the supply-side incentive of producing human content for model development. Second, in the content aggregation literature, the supply-chain relationship between the aggregator and content creators is one directional: content creators supply their work to aggregators. In our setting, such relationship is bi-directional: content creators may be customers to the AI firm (using the AI model to generate content), or suppliers (providing data for training), or occupy different roles over different periods. Finally, due to these distinctive technological and supply chain features discussed, we focus on the impact of two copyright policies -- fair use standard and copyrightability -- that do not arise in the content aggregation literature. Importantly, our research shows that due to the two-way interaction between the AI firm and content creators, these two copyright policies -- that are traditionally governed by different legal doctrines and examined independently -- are linked economically. In particular, we highlight that, when pre-existing training data are scarce, their interaction becomes so strong that they should be considered jointly.

Second, our paper is related to the emerging research stream focuses on the impact of generative AI on business, such as how the adoption of generative AI affects user productivity (e.g., \citealt{peng2023impact, noy2023experimental, dell2023navigating}) and creativity (e.g., \citealt{zhou2023generative}). Some of these papers focus on online labor market \citep{liu2023generate} and the content-creator and knowledge economy \citep{burtch2023consequences, huang2023generative, doshi2023generative}. Complementing this literature, our paper focuses on how various regulatory approaches related to copyright may shape the action and performance measures of different stakeholders within the creative industry. More broadly, our paper is related to the strand of research on how managers/agents make decisions in the presence of AI-assistance. Both empirical and theoretical works have examined this question in settings such as forecasting \citep{ibrahim2021eliciting}, retail \citep{kesavan2010inventory,sun2022predicting,caro2023believing}, sales \citep{luo2021artificial}, project management \citep{beer2022behavioral}, procurement \citep{cui2022ai}, adjudication \citep{cohen2023use}, car-sharing \citep{cui2024unlocking}, customer service \citep{xu2024identity}, and across different professions and occupations (e.g., \citealt{brynjolfsson2018can}, \citealt{liu2023ai}). Similar to the above research, our paper is also related to the implication of AI tools on business decisions (the data acquisition and model development decisions of AI companies, and the content production policy of creators). Differently, our paper focuses on the tradeoffs imposed by different regulatory decisions by capturing the interaction between the AI provider (and AI service) and users (creators and consumers). In addition to generating managerial insights, we also provide guidelines to policymakers to better understand the implications of different policy decision on various stakeholders and AI development.



Our paper is also related to the stream of literature that focuses on the impact of copyright protection on creative content. Several papers have empirically documented that copyright encourages creativity by increasing payment to authors \citep{li2018dead, giorcelli2020copyrights}. However, empirical research also finds that copyright  protection  reduces  consumers access to information goods and impedes the reuse of copyrighted materials in the production of new information  \citep{biasi2021effects, nagaraj2018does}. Our paper complements these literature by highlighting that the interaction between two copyright issues, copyrightability and the fair use standard, a salient feature in the context of generative AI. With the advent of generative AI, a burgeoning body of literature, mostly in law, addresses the complex legal challenges posed by this technology, in particular on the application of existing copyright doctrines to generative AI \citep{samuelson2023generative, epstein2023art, sag2023fairness}.  In a departure from these studies, through an analytical investigation into copyright issues in the context of generative AI, we unravel the potential ramifications of regulatory decisions by examining their impacts on the incentives and actions of various stakeholders. Finally, the work most closely related to ours is \cite{gans2024copyright}, who focuses on the issue of the fair use standard. Through an economic model featuring model size and negotiation cost, \cite{gans2024copyright} reveals that content creators should be compensated when the model is small, while the issue is more nuanced when the model is large. Our paper differs in two aspects: first, in addition to fair use, we also analyze the issue of AI-copyrightability and reveal the intricate link between these two important copyright issues. Second, instead of model size, our study unravels three other pivotal factors that significantly influence policy outcomes: the availability of human generated content as training data, the model quality and the competitive structure of the market.




\section{Model}\label{sect:model}
We consider a two-period model with three types of strategic players: a regulator, a (generative) AI company, and a continuum of content creators. An overview of the timeline is illustrated in Figure \ref{fig:timeline}, with details provided later.  In Period-1, the regulator first sets the two copyright-related decisions -- fair use standard and AI-copyrightability. The AI company, endowed with an AI model of quality $X_1$, then sets the price $p_1$ for each unit of AI usage. Given $X_1$ and $p_1$, each content creator decides whether to produce content through human creation, AI-assisted creation, or not to produce content at all. In Period 2, the AI company acquires training data, improves model quality from $X_1$ to $X_2$, and sets the second-period price $p_2$. In response, content creators make a new content-production decision.

\begin{figure}[htp]
	\begin{center}
		\caption{Model Timeline}
		\label{fig:timeline}
  \vspace{5mm}
		{
			\includegraphics[width=0.75\textwidth]{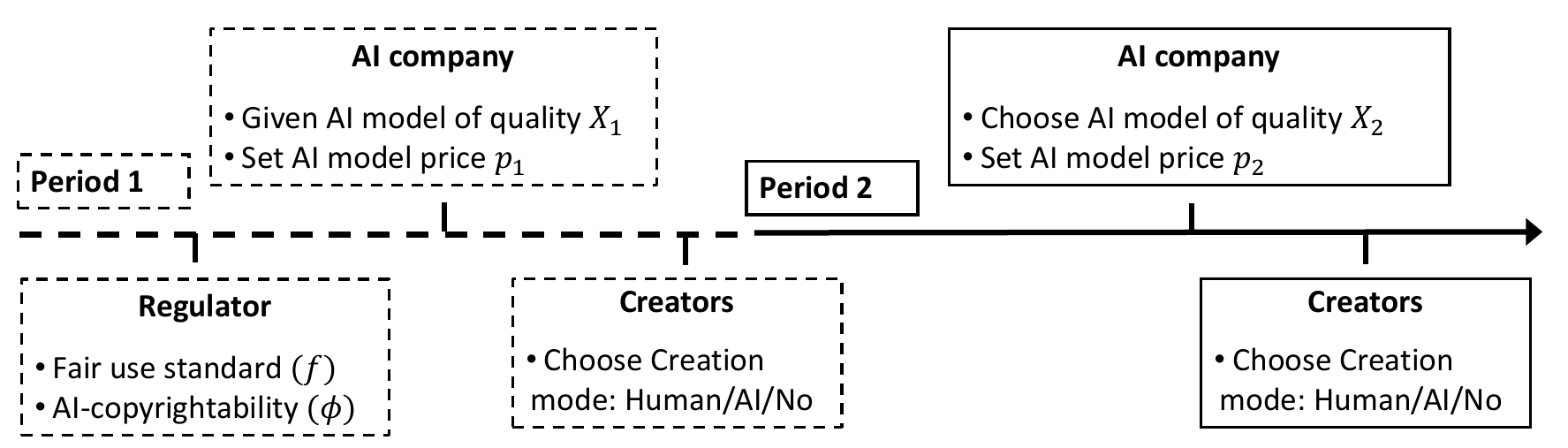}
			\label{fig:timeline}
		} 
	\end{center} 
\end{figure}

\subsection{Regulator}

At outset of Period-1, the regulator makes two decisions related to copyright that shape the regulatory environment of generative AI. 
\begin{enumerate}[(i)]
\item Fair use standard: the parameter $f \ge 0$ represents the amount of compensation the AI firm must pay content creators for each unit of content used in model training. We refer to the case with $f = 0$ (the AI company does not need to compensate copyright owners for training use) as {\em generous} fair use. Higher $f$ indicate a {\em stricter} fair use standard.
\item AI-copyrightability: the parameter $\phi \in [\underline{\phi}, 1]$ measures the degree to which AI-generated content receives copyright protection relative to human-created works. A higher value of $\phi$ indicates stronger protection; when $\phi = 1$, AI-generated content is treated identically to human content for copyright purposes. The lower bound $\underline{\phi}$ reflects that even without formal copyright protection, creators may still derive income from AI-generated works (for example, by selling them directly to publishers).
\end{enumerate} 
For the decision of AI-copyrightability to be relevant, we assume that human-created content and AI-generated one can be distinguished. While creators might try to pass off AI-generated content as human-created to enjoy stronger copyright protection (when $\phi < 1$), this risk can be managed. Jurisdictions such as China already require AI-generated content available to social media platforms to be clearly labeled, enabling provenance tracking \citep{feng2025china}. Similarly, California has introduced the AI Transparency Act, which establishes requirements for labeling and disclosing AI-generated content \citep{california2025AB853}. One could also impose substantial penalties for intentionally misrepresenting AI outputs as human-authored, treating such conduct as fraud. Combining disclosure requirements, deterrent sanctions, and improving detection tools makes it reasonable to assume that one can distinguish between human- and AI-generated content.

Finally, to capture the different regulatory priorities policymakers may pursue, we consider three policy objectives: (i) social welfare ($w$), defined as the sum of AI company profit, creator income, and consumer surplus; (ii) consumer surplus ($s$); and (iii) AI development, measured by Period-2 model quality ($X_2$). 

\subsection{Content Creators}
In each period ($i = 1, 2$), there is a continuum of content creators with mass $M_i$. To capture market growth, we assume $M_1 := 1 < M =: M_2$. Facing AI model quality $X_i$ and AI usage price $p_i$, each content creator -- who can produce one piece of content in each period -- chooses a content production policy. To maximize their payoffs (income minus cost), each creator chooses from three options: (1) do not create any content, which leads to a payoff of zero; (2) produce content through traditional means without using generative AI tools (human generation); and 3) produce content using the generative AI tool developed by the AI company. The latter two options differ in both both cost and income implications.

On the cost side, a content creator incurs a cost $c_H$ under human generation. Under AI generation, the cost consists of two components: the human cost of generation ($c_A$), and the price paid to the AI company for using AI usage ($p_i$, $i = 1, 2$). Correspondingly, the AI company earns revenue by providing this AI service to creators who choose AI generation. Specifically, the AI company's revenue at Period-$i$ is $p_i$ multiplied by the number of AI creators in this period. 

On the income side, a creator can earn revenue from two sources: (i) direct sales to consumers, and (ii) selling human-generated content to the AI firm for model-training purposes. We develop the second source in detail in the next subsection. For the first source, we assume that consumers' valuation of a work depends solely on its quality, not on the method of creation (human or AI).\footnote{We assume that, conditional on quality, consumers value human- and AI-generated content equally for expositional simplicity. Our structural results and qualitative insights remain unchanged if consumers value AI-generated content at some fraction of human-generated content.} A work can be either high quality (with valuation $v_h$) or low quality (with valuation $v_l$). For ease of exposition, we normalize $v_h = 1$ and $v_l = 0$. That is, only high-quality work has consumption value. Under copyright protection, a human-generated work can monetize a share $\beta \in (0, 1)$ of the consumption value, yielding an income of $\beta v_h = \beta$ if it is of high quality. By contrast, a high-quality AI-generated work yields $\phi \beta$, with $\phi$ being the level of AI-copyrightability. To rule out uninteresting cases, we impose the following assumption. 
\begin{assumption}
$c_A < c_H < \beta$. 
\end{assumption}
This assumption serves two purposes. First, $c_H < \beta$ ensures that, in the absence of AI tools, human creation is profitable for at least some content creators. Second, $c_A < c_H$ captures the cost saving of AI creation.

Content quality depends both on the creator's skill level and the mode of creation. Creators are endowed with heterogeneous skills in generating high-quality content, denoted by $x \in [0, 1]$,with cumulative distribution function $F()$. That is, a fraction of $F(x)$ of them have a skill level lower than or equal to $F(x)$. For tractability, we assume the skill levels of content creators are uniformly distributed, that is, $F(x) = x$. Under human generation, the probability that this piece of content is of high quality, denoted $x_H$, equals the creator's skill level $x$, the creator's skill. 

Under AI generation, by contrast, the probability of producing high-quality content depends on both the creator’s skill (e.g., conceptual input, prompt design) and the quality of the AI model. Let the model quality in Period-$i$ ($i = 1, 2$) be $X_i \in [0, 1]$. A creator with skill level $x$ produces high-quality content with probability 
\begin{align}
x_A = \lambda X_i + (1-\lambda) x,
\end{align}
where $\lambda \in [0,1]$ captures the degree of automation in content generation. Higher values of $\lambda$ imply less human involvement is required. This specification reflects the empirical findings that generative AI tools tend to improve the (relative) performance of low-skilled individuals more than the high-skilled ones \citep{dell2023navigating}.

\subsection{AI Company}
The AI company is assumed to operate in a competitive market. As such, its objective is to offer the most attract product to content creators subject to the constraint that its profit is non-negative. Without loss of generality, the AI company's marginal cost of providing AI service is normalized to zero. Consequently, in Period-1, because the AI firm incurs no cost, it always sets price $p_1 = 0$.

In Period-2, in addition to setting the price $p_2$, the AI company chooses how much data to acquire to improve its model. This decision depends on the cost of acquiring data, the supply of training data, and the impact of training data volume on model quality. As discussed above, the cost of data acquisition depends on the fair use standard: the AI company must pay a fee $f$ for each unit of content used in model training. This payment generates training income for creators and may influence their content-production choices. 

On the supply side, because AI-generated content has been shown to degrade model quality when used for training \citep{martinez2023towards, shumailov2023curse}, we assume that only high-quality human-generated content can be used to improve the model.\footnote{This assumption is made for expositional simplicity. Allowing AI-generated content to improve model quality -- albeit less effectively than human-generated content -- would not alter the core tradeoff or qualitative insights of this paper.} In Period-2, the AI company can source training data from two channels: pre-existing data not previously used for model training (with quantity $Q_0$) and high-quality human-generated content created during Period 1 ($Q_{1H}$). When the total amount of data the AI company chooses to acquire is smaller than the total supply ($Q_0 + Q_{1H}$), acquisition is randomly rationed across pre-existing data and newly created one.

Finally, to improve the model quality from $X_1$ to $X_2 \in [X_1, 1]$, the AI company must acquire the follow amount of data:
\begin{align}\label{eq:improving_X_2}
Q = q(X_2; X_1).
\end{align}
where the training data requirement function $q(\cdot)$ satisfies the following properties:
\begin{assumption}\label{ass:q}
$q(X_2; X_1)$ convexly increases in $X_2$, and decreases in $X_1$; $q(X_2; X_1) = 0$ for $X_2 \le X_1$, and 
\begin{align}
\lim_{X_2 \to 1} q(X_2; X_1) = \infty.    
\end{align} 
\end{assumption}
Intuitively, this assumption states that the amount of data needed to improve a model increases in the target model quality ($X_2$) and decreases in the initial quality ($X_1$). It also captures diminishing returns: progressively more data are required for incremental improvements, and reaching a “perfect” model ($X_2 = 1$) requires prohibitively large amounts of data. We adopt a general $q(\cdot)$ for all analytical results. For numerical examples in Section \ref{sect:policy}, we specify the training data requirement function as
\begin{align}
q(X_2; X_1) = \frac{1}{k} \log\left(\frac{1-X_1}{1-X_2}\right).    
\end{align} 
This functional form is motivated by the scaling laws in model training \citep{kaplan2020scaling,udandarao2024no}, which suggest that for the model quality to improve linearly, the required amount of training data grows exponentially. It is easy to verify that $q(X_2; X_1)$ satisfies all conditions in Assumption \ref{ass:q}. The parameter $k$ captures {\em training efficiency}: a larger $k$ implies that less data are required to achieve a given level of model improvement.

\section{Equilibrium Analysis}\label{sect:equilibrium}
Building on the model above, this section analyzes the equilibrium decisions of the AI company and content creators, deferring discussion of the effects of copyright policy to \S\ref{sect:policy}.

\subsection{Period-2 Content Creation}\label{sect:period_2}
We solve the game in backward induction, beginning with content creators' production policy in Period-2. Here, given an AI model with quality $X_2$ and the AI usage price $p_2$, each creator chooses among three options: produce human-generated content, produce AI-generated content, or produce no content. For human creation, the expected payoff for a creator with skill level $x$ is:
\begin{align}\label{eq:u_H}
u_H = \beta x - c_H.  
\end{align}
If the creator instead chooses AI generation, the expected payoff is
\begin{align}\label{eq:u_A}
u_A = \phi \beta [\lambda X_2 + (1-\lambda)x] - c_A - p_2,    
\end{align}
where $X_2$ is the improved AI model quality, and $\phi$ captures the level of AI-copyrightability. Finally, if a creator decides not to produce any content, their payoff is $u_N = 0$. By comparing these expected payoffs, each creator select the option that maximizes their utility.

To characterize the creators' Period-2 policy, we define the following quantities.
\begin{align}\label{eq:threshold_model}
X_A(p) = \frac{c_A + p - (1-\lambda)\phi c_H}{\lambda \phi \beta}; \quad X_H(p) = \frac{c_A + p + [1-(1-\lambda)\phi]\beta - c_H}{\lambda \phi \beta}; \quad X_N(p) = \frac{c_A + p}{\lambda \phi \beta}.
\end{align}
and
\begin{align}\label{eq:threshold_skill}
x_{HO} = \frac{c_H}{\beta}; \quad x_{HA}(X, p) = \frac{\lambda \phi \beta X + (c_H - c_A - p) }{[1 - (1-\lambda) \phi]\beta}; \quad x_{AO}(X, p) = \frac{c_A + p - \lambda \phi \beta X}{(1-\lambda) \phi \beta}.
\end{align}
We further define functions $\underline{x}(X, p)$ and $\overline{x}(X, p)$ in Table \ref{table:second_period_threshold}.
\begin{table}[htp]
\renewcommand{\arraystretch}{1.3}
    \centering
    \caption{Thresholds for Second Period Production Policy}
    \label{table:second_period_threshold}
    \vspace{2mm}
    \begin{tabular}{ccccc}
        \multicolumn{5}{c}{Case I: $\phi \le \frac{\beta - c_H}{\beta(1-\lambda)}$ }\\
        \hline
        $X$ & $[0, X_A(p)]$ & $[X_A(p), X_N(p)]$  & $[X_N(p), X_H(p)]$ & $[X_H(p),  1]$\\
        $(\underline{x}(X,p), \overline{x}(X,p))$ & $(x_{HO}, x_{HO})$ & $(x_{AO}(X,p), x_{HA}(X,p))$ & $(0, x_{HA}(X,p))$ & $(0, 1)$ \\
\hline \\
         \multicolumn{5}{c}{Case II: $\phi > \frac{\beta - c_H}{\beta(1-\lambda)}$ }\\
         \hline
         $X$ & $[0, X_A(p)]$ & $[X_A(p), X_H(p)]$  & $[X_H(p), X_N(p)]$ & $[X_N(p),  1]$\\
         $(\underline{x}(X,p), \overline{x}(X,p))$ & $(x_{HO}, x_{HO})$ & $(x_{AO}(X,p), x_{HA}(X,p))$ & $(x_{AO}(X,p), 1)$ & $(0, 1)$ \\
        \hline
    \end{tabular}
  \end{table}

\begin{proposition}\label{prop:second_period_production}
In Period-2, given AI model quality $X_2$ and AI usage price $p_2$, creators' content-production policy is:
\begin{enumerate}
    \item creators with skill level $x < \underline{x}(X_2, p_2)$ do not produce any content;
    \item creators with $x \in [\underline{x}(X_2, p_2), \; \overline{x}(X_2, p_2))$ produce AI content; and
    \item creators with $x \ge \overline{x}(X_2, p_2)$ produce human content. 
\end{enumerate}
\end{proposition}
Proposition \ref{prop:second_period_production} shows that creators follow a double-threshold production policy. Those with the lowest skill levels find it unprofitable to produce any content, while those with the highest skill levels choose to produce content without using AI. Creators with intermediate skill levels opt for AI generation. The size of these three ranges depends on the AI model quality ($X_2$) and the AI service price ($p_2$): as the AI tool becomes more capable (or the tool becomes cheaper), the range of AI creation expands, whereas the two regions shrink. In other words, a stronger AI model incentivizes more AI creation both by substituting for human creation and by inducing new entry. Lower AI usage prices have a similar effects. Finally, the thresholds depend on AI-copyrightability ($\phi$), highlighting the direct influence of copyright policy on creators’ production choices.


\begin{corollary}\label{coro:second_period_performance}
In Period-2, the quantities of (high-quality) human content ($Q_{2H}$) and (high-quality) AI content ($Q_{2A}$) generated are given by:
\begin{align}
Q_{2H} = \frac{M(1-\overline{x}^2)}{2}, \quad Q_{2A} = M \lambda X_2 (\overline{x} - \underline{x}) + \frac{M(1-\lambda)}{2}\left(\overline{x}^2 - \underline{x}^2\right).    
\end{align}
The AI company's revenue ($r_2$), the content creators' aggregated income ($u_2$), consumer surplus ($s_2$), and social welfare ($w_2$) are:
\begin{align}
r_2 &= M p_2 \left(\overline{x} - \underline{x}\right); \\
u_2 &= \beta (Q_{2H} + \phi Q_{2A}) - M \left[c_H (1-\overline{x}) + (c_A + p_2) \left(\overline{x} - \underline{x}\right) \right]; \\
s_2 &= (1-\beta)Q_{2H} + (1-\phi \beta)Q_{2A}. \\
w_2 &= Q_{2H} + Q_{2A} - M \left[c_H (1-\overline{x}) + c_A (\overline{x} - \underline{x}) \right]. \label{eq:w_2}
\end{align}
Moreover, $Q_{2A}$ and $u_2$ increase in $X_2$. $r_2$ increases concavely in $X_2$ when $X_2 \ge X_A(p)$, and for $X_2 \ge X_A(0)$, $r_2$ is concave in $p_2 \in [0, \overline{p}(X_2)]$, where $\overline{p}(X) = \lambda \phi \beta X + (1-\lambda)\phi c_H - c_A$.
\end{corollary}
Corollary \ref{coro:second_period_performance} illustrates how AI model quality shapes the volume of content created and the payoffs of different stakeholders. As the AI model improves, more AI-generated content is produced and creators’ aggregate income rises. The AI company’s revenue also increases because stronger models attract more creators to adopt AI generation. Further, when the model quality is sufficiently high that AI creation is relevant ($X_2 \ge X_A(0)$), the firm’s revenue $r_2$ is concave in $p_2$ over the range in which AI creation occurs ($\overline{x} > \underline{x}$). Since $r_2 = 0$ at both $p_2 = 0$ and $p_2 = \overline{p}(X_2)$, revenue first increases and then decreases in $p_2$.

Finally, note that Period-2 social welfare ($w_2$) is the sum of AI company revenue ($r_2$), creator income ($u_2$), and consumer surplus ($s_2$). It is independent of the AI model training cost that AI company incurs under strict fair use ($f > 0$). This is because such model training cost ($fq(X_2, X_1)$) is a transfer payment from the AI company to content creators (either those in Period-1, or those providing pre-existing training content), and therefore does not affect total welfare.

\subsection{Period-2 Price Setting and AI Model Improvement}\label{sect:model_development}
Anticipating how its revenue in Period-2 ($r_2$) is jointly affected by AI service price ($p_2$) and model quality ($X_2$), the AI company chooses $p_2$ and the amount of data to acquire for model training, which in turn determines $X_2$. Because the AI company operates in a competitive market, it selects $p_2$ and $X_2$ to maximize the value of its AI service to AI content creators, subject to the constraint that its profit  -- $\pi_2(X_2, p_2)$ -- is non-negative. By joining its revenue and model training cost, the company's profit is:
\begin{align}\label{eq:profit_2nd_period}
\pi_2(X_2, p_2) = r_2(X_2, p_2) - f q(X_2; X_1) = M p_2 \left[\overline{x}(X_2, p_2) - \underline{x}(X_2, p_2)\right] - f q(X_2; X_1),
\end{align}
where the first term is the firm's revenue under model quality $X_2$ and price $p_2$, and the second term represents the cost of acquiring training data, using the one-to-one correspondence between $Q$ and $X_2$ (Eq. \ref{eq:improving_X_2}). Under the decision $(X_2, p_2)$, the value of the AI service for a content creator follows Eq.~\eqref{eq:u_A}. Combined, the company's optimization problem can be formalized as:
\begin{align}
&\max_{X_2, p_2} \;\; \phi \beta \lambda X_2 - p_2; \\
& \mbox{s.t.} \;\;  \pi_2(X_2, p_2) \ge 0.
\end{align}
Based on this formulation, it is clear that because $X_2$ always increases the firm's revenue. Therefore, under generous fair use, as data acquisition does not incur any cost, the AI company will acquire as much data as possible to improve the AI model, leading to the following result. 
\begin{proposition}\label{prop:model_improvement_generous}
Under generous fair use standard ($f = 0$), the AI service price is $p_2^* = 0$. The improved model quality is:
\begin{align}\label{eq:X_2_G}
X_2^* = X_2^G := q^{-1}(Q_0 + Q_{1H}^G; X_1),
\end{align}
where $Q_{1H}^G$ denotes the total amount of high-quality human content created in Period-1 under generous fair use. 
\end{proposition}
Proposition \ref{prop:model_improvement_generous} implies that if pre-existing training data are abundant ($Q_0 \to +\infty$), then $X_2^G = 1$, independent of Period-1 content creation. By contrast, when pre-existing data are limited, the AI company uses all high-quality human content created in Period 1 ($Q_{1H}^G$). As we show later, this means that the amount of human content produced in Period 1 can become a binding constraint on Period-2 model quality.

On the other hand, when the fair use standard is strict ($f > 0$), the firm's optimization problem becomes more intricate. Similar to the generous fair use case, the improved model quality is also bounded by data availability, that is, $X_2 \le X_2^D := q^{-1}(Q_0+Q_{1H}^S; X_1)$, where $Q_{1H}^S$ denotes the amount of human content created in Period-1. We note that, while the bound shares the same form as in the generous fair use case, the amounts of Period-1 human created data ($Q_{1H}^S$ vs. $Q_{1H}^G$) may be different. In addition, as $q'()$ approaches infinity as $X_2$ approaches 1, model training cost -- $f q(X_2; X_1)$ -- eventually dominates the revenue. Thus, we only need to consider those $X_2$ such that $\pi_2(X_2, p) \ge 0$ for at least some $p \ge 0$, or more formally, $X_2 \in \Omega_2(X_1)$ where
\begin{align}
\Omega_2 := \{X \in [X_1, X_2^D] \; | \; \max_{p \ge 0} \pi_2(X,p) \ge 0\}.
\end{align}
Let $p_2^X(X_2) = \min_{p} \{\pi_2(X_2, p) \ge 0\}$ for $X_2 \in \Omega_2(X_1)$, which represents the AI service price the company will set to break even given model quality $X_2$ under competitive pressure.
\begin{lemma}\label{lem:model_improvement_strict}
Under strict fair use ($f > 0$), the improved model quality $X_2^*$ is:
\begin{align}
X_2^* = \arg\max_{X_2 \in \Omega_2} \; v_A(X_2) := \phi \beta \lambda X_2 - p_2^X(X_2),
\end{align}
and the price is $p_2^* = p_2^X(X_2^*)$. The total amount of training data required for model improvement is $Q^A = q(X_2^*; X_1)$.
\end{lemma}
The following proposition provides the structure of the AI company's equilibrium choice. Let 
\begin{align}
\overline{X}_2 &= \max\{X_2: \max_{p_2 \ge 0} \; \pi_2(X_2, p_2) \ge 0\}.
\end{align}

\begin{proposition}\label{prop:model_improvement_strict_large_X}
Under strict fair use ($f > 0$) and $X_1 \ge X_A(0)$, 
\begin{enumerate}[(i)]
\item $\Omega_2 = [X_1, \min(\overline{X}_2, X_2^G)]$;
\item for $X_2 \in \Omega_2(X_1)$, $p_2^X$ increases in $X_2$;
\item if $q''(\cdot)$ is sufficiently large, $p_2^X(X_2)$ is convex on $X_2$, and 
\begin{enumerate}
\item if $\phi \beta \lambda \le \frac{\partial p_2^X}{\partial X_2}|_{X_2=X_1}$, the firm sets $X_2^* = X_1$, $p_2^* = 0$;
\item if $\phi \beta \lambda \in \left(\frac{\partial p_2^X}{\partial X_2}|_{X_2=X_1}, \; \frac{\partial p_2^X}{\partial X_2}|_{X_2=\min(\overline{X}_2, X_2^D)} \right)$, $X_2^* \in (X_1, \min(\overline{X}_2, X_2^D))$ is uniquely determined by $\frac{\partial p_2^X}{\partial X_2}|_{X_2=X_2^*} = \lambda \phi \beta$.
\item if $\phi \beta \lambda \ge \frac{\partial p_2^X}{\partial X_2}|_{X_2=\min(\overline{X}_2, X_2^D)}$, the firm sets $X_2^* = \min(\overline{X}_2, X_2^D)$ and $p_2^* = p_2^X(X_2^*)$.
\end{enumerate}
\end{enumerate}
\end{proposition}
As shown, when the initial model quality is high -- so that even some creators will adopt AI creation without any Period-2 model improvement ($X_2 = X_1$) -- the feasible set of $X_2$ is $[0, \min(X_2^D, \overline{X}_2)]$. Here $X_2^D$ is determined by training data availability, while $\overline{X}_2$ is governed by training cost. Within this interval, the firm increases the corresponding service price ($p_2^X$) as model training becomes increasingly costly. If the model training cost is sufficiently convex, the required price also increases convexly. In this case, the optimal $(X_2, p_2)$ is determined by the the relative magnitude of the marginal benefit of model improvement for AI creators ($\lambda \phi \beta$), and the marginal cost ($\frac{\partial p_2^X}{\partial X_2}$). When the marginal cost already exceeds the marginal benefit at $X_2 = X_1$, it is optimal for the AI company to forgo model improvement, and set $p_2^* = 0$ (Scenario 1). Conversely, if the marginal cost is sufficiently sufficiently low, the AI company should improve the model to the highest possible level, bounded either by $\overline{X}_2$ or $X_2^D$, as in Scenario 3. Intermediate cases yield an interior optimum in which the firm uses only part of the available data for training.

The proposition also reveals how AI firm's Period-2 decisions are affected by copyright policy $(f, \phi)$.The marginal benefit of improvement ($\lambda \phi \beta$) depends directly on AI-copyrightability. In addition, note that
\begin{align}
\frac{\partial p_2^X}{\partial X_2} = \frac{f \frac{\partial q}{\partial X_2} - \frac{\partial r_2}{\partial X_2}}{\frac{\partial r_2}{\partial p_2}}.
\end{align}
By the discussion of Corollary \ref{coro:second_period_performance}, the marginal cost ($\frac{\partial p_2^X}{\partial X_2}$) is influenced by $\phi$ through both $\frac{\partial r_2}{\partial X_2}$ and $\frac{\partial r_2}{\partial p_2}$. The expression also shows how the fair use standard ($f$) affects marginal cost: an overly strict fair-use policy raises $\frac{\partial p_2^X}{\partial X_2}$ and can ultimately hinder AI development.

\begin{proposition}\label{prop:model_improvement_strict_small_X}
Under strict fair use ($f > 0$) and $X_1 < X_A(0)$, 
\begin{enumerate}
\item there exists $\underline{X}_2 > X_1$ such that $\Omega = \{X_1\} \cup [\underline{X}_2, \min(X_2^G, \overline{X}_2)]$;
\item if $Q_0 + Q_{1H}^S$ is sufficiently small, the firm sets $X_2^* = X_1$ and $p_2^* = 0$.
\end{enumerate}
\end{proposition}
Intuitively, when improved model quality lies below the threshold $X_A(0)$, no creator adopts AI generation in Period 2. Consequently, the firm earns zero revenue in this region, and any additional model improvement only increases data-acquisition cost, reducing profit. Therefore, the AI firm will never improve the model quality only marginally. Further, if the available training data $(Q_0 + Q_{1H})$ are limited, it is optimal for the AI company to forgo model improvement altogether and set $p_2^* = 0$.

\subsection{Period-1 Content Creation}
Returning to the Period-1, recall from Section \ref{sect:model}, because the AI company incurs no cost in this period, competitive pressure drives the AI service price to $p_1^* = 0$.
\begin{proposition}\label{prop:first_period_abundant}
Under abundant pre-existing training data ($Q_0 \to +\infty$) or generous fair use ($f = 0$), content creator's Period-1 policy under the AI model quality $X_1$ and AI service price $p_1^* = 0$ is
\begin{enumerate}
    \item creators with skill level $x < \underline{x}(X_1, 0)$ do not produce any content;
    \item creators with $x \in [\underline{x}(X_1, 0), \; \overline{x}(X_1, 0))$ produce AI content; and
    \item creators with $x \ge \overline{x}(X_1, 0)$ produce human content,
\end{enumerate}
where $\overline{x}(\cdot,\cdot)$ and $\underline{x}(\cdot,\cdot)$ are defined in Table \ref{table:second_period_threshold}. 
\end{proposition} 
Intuitively, under generous fair use standard, creators receive no compensation for selling human content they generated in Period-1, so their creation policy is the same as that in Period-2 (Proposition \ref{prop:second_period_production}), except that thresholds $\overline{x}$ and $\underline{x}$ are now based on the initial AI model quality is $X_1$ and price ($p_1^* = 0$). Similarly, when pre-existing training data are abundant ($Q_0 \to \infty$), income from selling human content they generated in Period-1 becomes negligible because the AI company primarily relies on pre-existing data under random rationing. Under either condition, the Period-1 performance measures take the same form as those in Period-2 (Corollary \ref{coro:second_period_performance}), except that the creator mass is $1$ instead of $M$.

However, with strict fair use ($f > 0$) and limited pre-existing data, creators' Period-1 production decisions become more complex. When producing human content, can earn not only from consumption value but also potentially from training value ($f$), with the income from the latter channel affected by the collective production decisions by all creators. For example, suppose that $f$ is relatively small. In this case, the AI company will seek to acquire a substantial amount of human data for training. Let this desired amount be $Q^C$. When $Q^C \gg Q_0$, all human content created in Period-1 will be acquired at price $f$. A creator’s payoff from human generation is then
\begin{align}\label{eq:u_H_F}
u_H^F(x) = (\beta + f) x - c_H.
\end{align}
Compared to Eq. \eqref{eq:u_H}, the newly added term ($fx$) captures the extra incentive for creators to produce human content due to training income. At the other extreme, when $f$ is very large, the AI company the AI company finds model improvement too costly and chooses $Q^C = 0$. Accordingly, human creation yields no training income, and creators' payoff and and creators’ Period-1 decisions revert to the generous fair use case (Proposition \ref{prop:first_period_abundant}). For intermediate $f$, if the total human content available -- both pre-existing ($Q_0$) and created in Period-1 ($Q_{1H}$) -- exceeds $Q^C$, each piece of human content is acquired for training with probability $\frac{Q^C}{Q_0 + Q_{1H}}$. Anticipating that, a creator’s expected payoff from human generation becomes
\begin{align}\label{eq:u_H_P}
u_H^P(x) = \left[\beta + \left(\frac{Q^C}{Q_0 + Q_{1H}}\right)f \right] x - c_H.
\end{align}
As shown, the creator's payoff depends not only on $Q^C$, which reflects the AI company's demand for data, but also on $Q_{1H}$, the aggregate supply of (high-quality) human content created in Period 1. This interaction generates a self-regulating mechanism. When the AI company’s expected demand for training data rises (e.g., due to stronger revenue projections in Period 2), creators anticipate higher training income and are more likely to produce human content. Conversely, as more creators switch to human generation and $Q_{1H}$ increases, the likelihood of any single piece being acquired decreases, reducing expected training income. Considering all three scenarios, we obtain the following characterization of creators’ Period-1 production policy, with thresholds $(\underline{x}^S, \overline{x}^S)$ defined in Appendix \ref{appx:first_period_thresholds} for expositional brevity.
\begin{proposition}\label{prop:first_period_strict}
Under strict fair use ($f > 0$) and finite pre-existing data ($Q_0$), creators' Period-1 content production policy is:
\begin{enumerate}
\item creators with skill level $x < \underline{x}^S$ do not produce any content;
\item creators with $x \in (\underline{x}^S, \overline{x}^S]$ produce AI content; and 
\item creators with $x > \overline{x}^S$ produce human content.
\end{enumerate}
Under this policy, the amount of (high-quality) human-generated content is:
\begin{align}
Q_{1H}^S = \frac{1 - (\overline{x}^S)^2}{2}.
\end{align}
\end{proposition}
Thus, despite the additional dynamics introduced by training incentives, creators’ production structure remains qualitatively similar to Proposition \ref{prop:second_period_production}. However, as implied by Eqs.~\eqref{eq:u_H_F} and \eqref{eq:u_H_P}, the relevant thresholds $\underline{x}^S$ and $\overline{x}^S$ now depend on additional factors such as the fair-use parameter ($f$), market size ($M$), and model-training efficiency (captured by $q(\cdot)$). This result highlights an important intertemporal linkage between creators’ Period-1 decisions and the AI company’s model-improvement strategy. As we show later, this linkage plays a crucial role in determining the effectiveness of copyright regulation.

\section{Impact of Copyright Policies}\label{sect:policy}
With the above equilibrium results, we now examine how copyright policies affect AI development (as measured by improved model quality $X_2$) and performance outcomes for different stakeholders, including creator income, consumer surplus, and social welfare. To isolate the distinct channels through which copyright policies shape decisions and outcomes, we focus on two special cases defined by the availability of pre-existing training data ($Q_0$):
\begin{enumerate}[(i)]
\item The data {\em abundant} regime. This regime corresponds to a setting with abundant pre-existing data for model training ($Q_0 \to +\infty$). It captures scenarios in which model development is at an early stage (e.g., video-generation models) and substantial amounts of legacy data remain unused in training the initial model ($X_1$). In this case, the AI company does not rely on content newly created in the current period.
\item The data {\em scarce} regime. This regime corresponds to a setting with no pre-existing content available for model training ($Q_0 = 0$). It reflects environments in which the model is already highly developed and prior data have largely been exhausted (e.g., text generation models), or where existing data become obsolete for technological or legal reasons. Here, the AI company relies solely on high-quality human-generated content from Period 1 ($Q_{1H}$).
\end{enumerate}

\subsection{Copyright Policy in the Data Abundant Regime}\label{sect:policy_abundant}
In the presence of abundant pre-existing training data, Proposition \ref{prop:first_period_abundant} shows that Period-1 content-production decisions and Period-2 model-improvement decisions become independent. This separation allows us to isolate the impact of copyright regulation without intertemporal feedback. We begin by examining the two policy instruments -- fair-use standard and AI-copyrightability -- individually, and then study their interaction. We conclude by exploring how market conditions, technological characteristics, and regulatory objectives shape the choice of copyright policy.

\subsubsection{Fair Use Standard.}
\begin{figure}[htp]
	\begin{center}
		\caption{Impact of fair use standard under the data abundant regime: consumer surplus}
		\label{fig:abundant_fair_use}
  \vspace{5mm}
	\subfloat[Current status of AI ($X_1$)]
		{
			\includegraphics[width=0.3\textwidth]{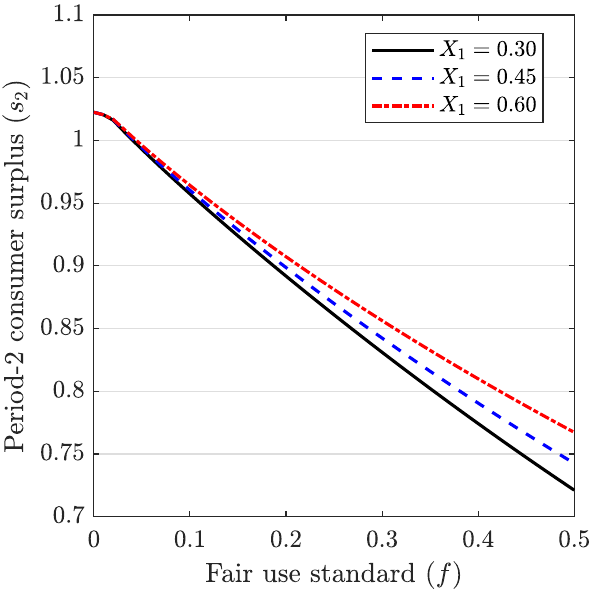}
			\label{fig:abundant_fair_use_X_1}
		} \quad 
		\subfloat[Training efficiency ($k$)]
		{
			\includegraphics[width=0.3\textwidth]{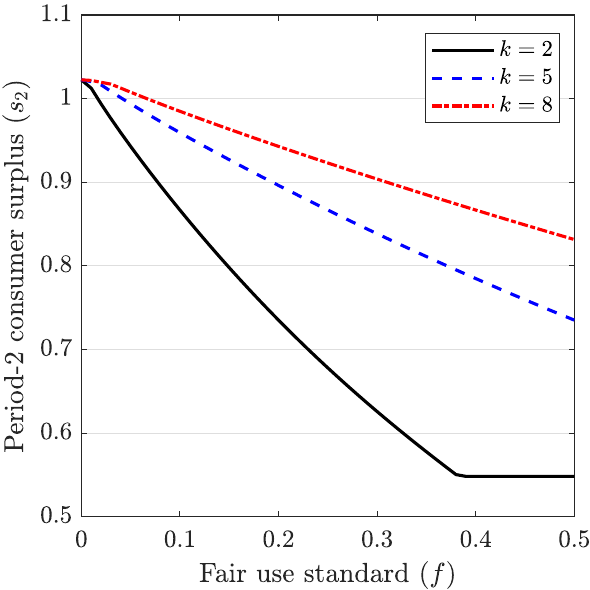}
			\label{fig:abundant_fair_use_k}
		} \quad
        \subfloat[Market growth ($M$)]
		{
			\includegraphics[width=0.3\textwidth]{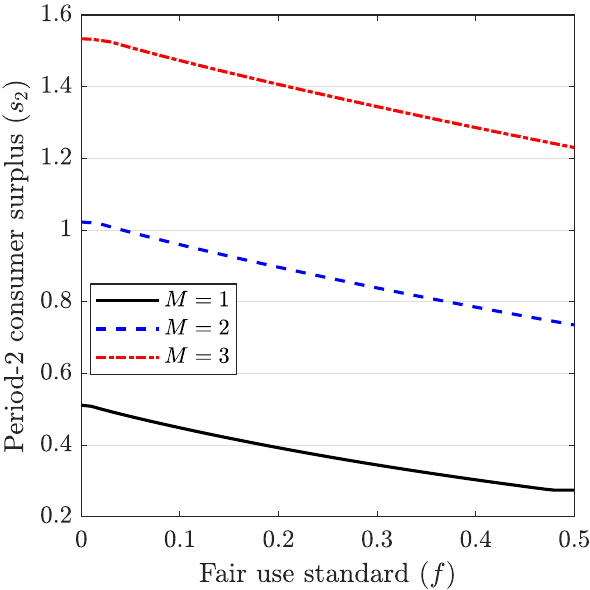}
			\label{fig:abundant_fair_use_M}
		} \\ 
	\end{center}
	\vspace{2mm}
	{\footnotesize \noindent
	\underline{\textbf{Parameters}}: $c_H = 0.3$, $c_A = 0.05$, $\lambda = 0.6$, $\beta =0.6$; $\phi = 0.6$; $M = 2$ for (a) and (b); $X_1 = 0.4$ for (b) and (c); $k = 5$ for (a) and (c).
	} 
\end{figure}

As established in Proposition \ref{prop:first_period_abundant}, with abundant pre-existing data, fair use standard has no impact on creators' Period-1 decisions. Thus, its role in this regime is direct: by affecting the cost of inputs to the AI model, it influences the AI company's model development decision ($X_2$), and consequently, all Period-2 performance outcomes.
\begin{proposition}\label{prop:abundant_fair_use}
In the data abundant regime, generous fair use ($f = 0$) maximizes model quality ($X_2$), the amount of (high-quality) AI content, creator income ($u_2$) in Period-2, and overall social welfare ($w_2$). 
\end{proposition}
Further, as illustrated in Figure \ref{fig:abundant_fair_use}, consumer surplus is also maximized under generous fair use across a range of technological and market conditions, including initial model quality ($X_1$, Figure \ref{fig:abundant_fair_use_X_1}), training efficiency ($k$, Figure \ref{fig:abundant_fair_use_k}), and market growth ($M$, Figure \ref{fig:abundant_fair_use_M}). Taken together, when training data are plentiful, lowering AI training costs through generous fair use benefits all major stakeholders.

\subsubsection{AI-Copyrightability.}
Unlike fair use standard, AI-copyrightability influences decisions in both periods. The following proposition characterizes its impact in Period-1.

\begin{figure}[htp]
	\begin{center}
		\caption{Impact of AI-copyrightability under the data abundant regime: Period-1 outcomes}
		\label{fig:abundant_copyrightability_1}
  \vspace{5mm}
		\subfloat[Consumer surplus ($s_1$)]
		{
			\includegraphics[width=0.3\textwidth]{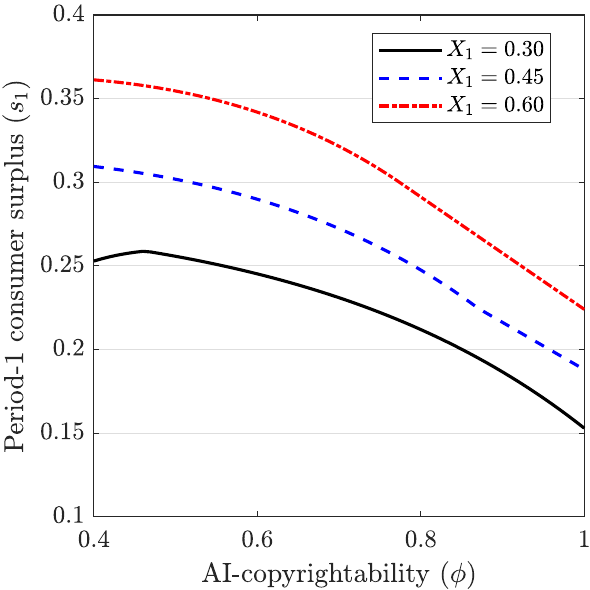}
			\label{fig:abundant_copyrightability_s_1}
		} \quad \quad
		\subfloat[Social welfare ($w_1$)]
		{
			\includegraphics[width=0.3\textwidth]{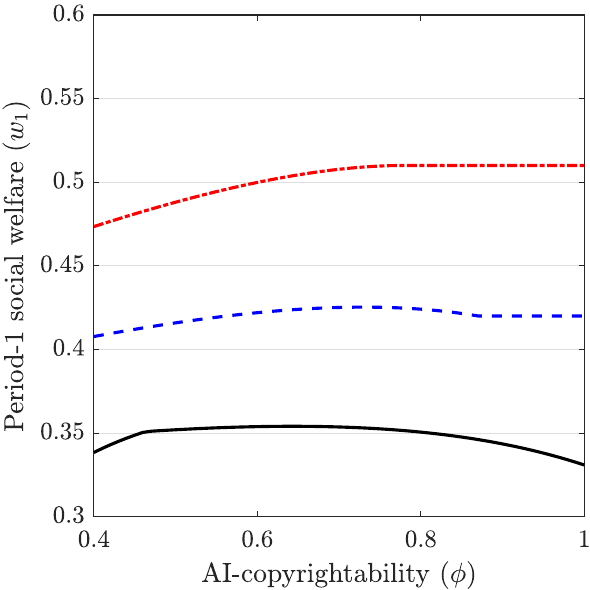}
			\label{fig:abundant_copyrightability_w_1}
		} \\
	\end{center}
	\vspace{2mm}
	{\footnotesize \noindent
	\underline{\textbf{Parameters}}: $c_H = 0.3$, $c_A = 0.05$, $\lambda = 0.6$, $\beta =0.6$, $M = 2$, $k = 5$, $f = 0.2$.
	} 
\end{figure}

\begin{proposition}\label{prop:abundant_copyrightability_1}
In the data abundant regime, both the amount of Period-1 AI content and Period-1 creator income increase in AI-copyrightability ($\phi$). When $X_1$ is sufficiently large, Period-1 social welfare also increases in $\phi$. 
\end{proposition}
The result highlights the direct effect of AI-coyprightability: by making the AI creation optimal more attractive financially, a higher $\phi$ induces more creators to adopt AI tools and increases aggregate creator income. Its impact on consumers, however, is more nuanced. On the positive side, strong AI-copyrightability benefits consumers by attracting more creators to enter the market and encouraging existing AI creators to produce more high-quality AI content. On the negative side, strong AI-copyrightability allocates a larger share of the consumption value to the creators, reducing the portion that accrues to consumers. Moreover, higher $\phi$ may incentivize creators to switch from producing higher-quality human content to lower-quality, but cheaper-to-produce, AI content, potentially harming consumers. As illustrated in Figure \ref{fig:abundant_copyrightability_s_1}, these negative channels can dominate, causing consumer surplus to fall as $\phi$ increases. The pattern for social welfare, shown in Figure \ref{fig:abundant_copyrightability_w_1}, reflects Proposition \ref{prop:abundant_copyrightability_1}: when initial model quality is high, switching from human to AI generation primarily lowers production costs without substantially reducing quality, and stronger AI-copyrightability improves social welfare. By contrast, when $X_1$ is low, increases in $\phi$ induce creators to adopt AI generation mainly to reduce costs, which degrades content quality and harms social welfare.

\begin{figure}[htp]
	\begin{center}
     \caption{Impact of AI-copyrightability under data abundant regime: Period-2 outcomes}\label{fig:abundant_copyrightability_2}	
  \vspace{5mm}
		\subfloat[AI development ($X_2$)]
		{
			\includegraphics[width=0.3\textwidth]{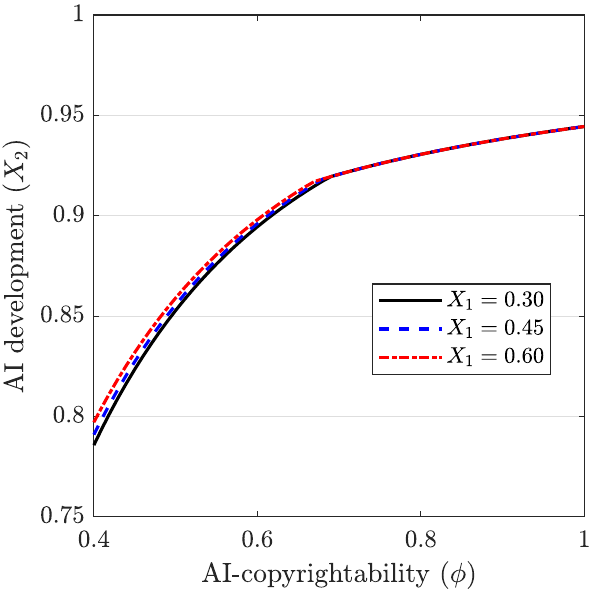}
			\label{fig:abundant_copyrightability_X_2}
		} \quad 
        \subfloat[{Consumer surplus ($s_2$)}]
		{
			\includegraphics[width=0.3\textwidth]{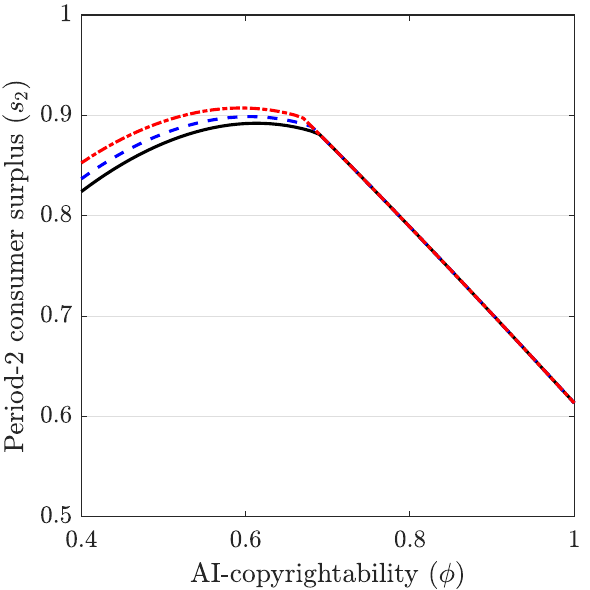}
			\label{fig:abundant_copyrightability_s_2}
		} \quad
		\subfloat[{Social welfare ($w_2$)}]
		{
			\includegraphics[width=0.3\textwidth]{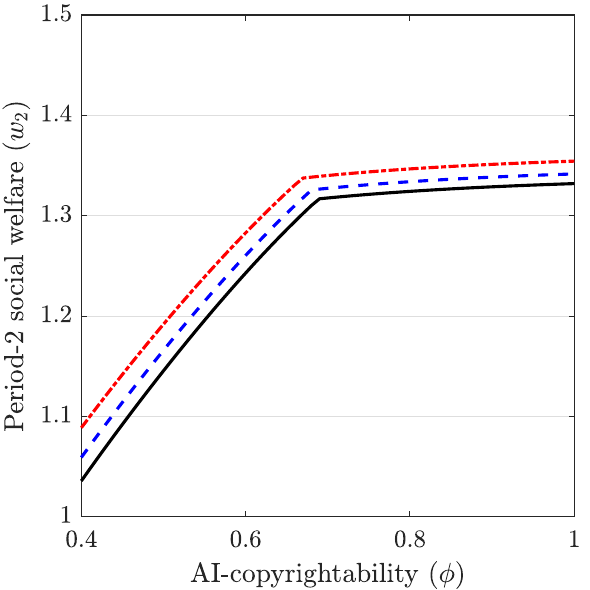}
			\label{fig:abundant_copyrightability_w_2}
		} \\ 
	\end{center}
	\vspace{2mm}
	{\footnotesize \noindent \underline{\textbf{Parameters}}: $c_H = 0.3$, $c_A = 0.05$, $\lambda = 0.6$, $\beta =0.6$, $M = 2$, $k = 5$; $f = 0.2$.}
\end{figure}

In Period 2, beyond the direct effect described above, AI-copyrightability also exerts an indirect influence by shaping the AI company’s model-development decision ($X_2$). Combining these two channels, the following proposition shows that stronger AI-copyrightability is always beneficial to content creators.
\begin{proposition}\label{prop:abundant_copyrightability_2}
In the data abundant regime, the Period-2 creator income increases in AI-copyrightability ($\phi$).
\end{proposition}
Figure \ref{fig:abundant_copyrightability_X_2} illustrates the impact of AI-copyrightability on model development. As AI-generated content receives greater copyright protection, demand for the AI company’s services rises, which incentivizes further improvements in model quality. This additional effect generally makes AI-copyrightability more favorable for consumer surplus and social welfare, as evidenced by comparing Figure \ref{fig:abundant_copyrightability_s_1} (resp. Figure \ref{fig:abundant_copyrightability_w_1}) with Figure \ref{fig:abundant_copyrightability_s_2} (resp. Figure \ref{fig:abundant_copyrightability_w_2}).

\subsubsection{Interaction of Fair Use Standard and AI-Copyrightability.}
\begin{figure}[htp]
	\begin{center}
		\caption{Interaction of fair use standard and AI-copyrightability under the data abundant regime}
		\label{fig:abundant_interaction}

        \vspace{3mm}
        \subfloat[AI development ($X_2$)]
		{
			\includegraphics[width=0.3\textwidth]{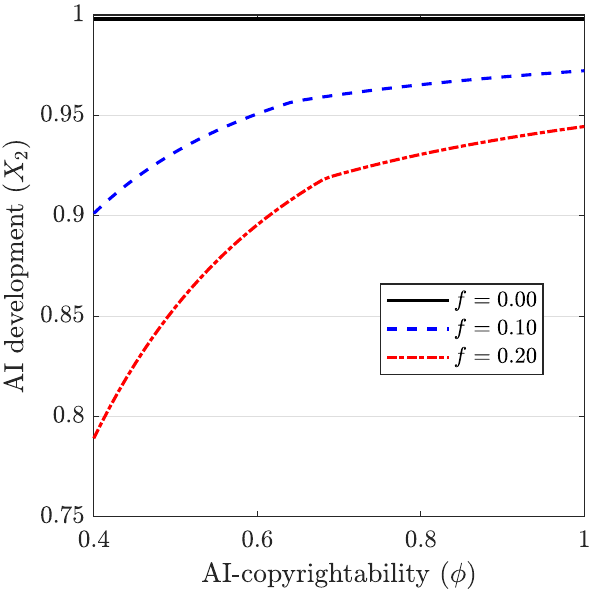}
			\label{fig:abundant_interaction_X_2}
		} \quad
		\subfloat[Long-term consumer surplus ($s$)]
		{
			\includegraphics[width=0.3\textwidth]{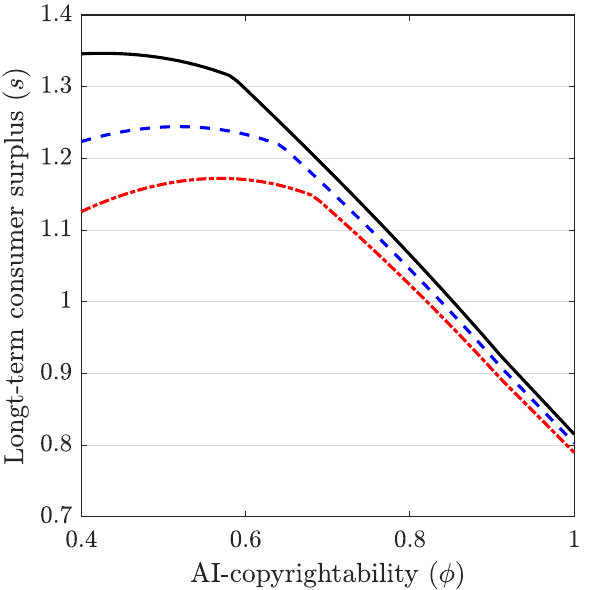}
			\label{fig:abundant_interaction_s}
		} \quad
		\subfloat[Long-term social Welfare ($w$)]
		{
			\includegraphics[width=0.3\textwidth]{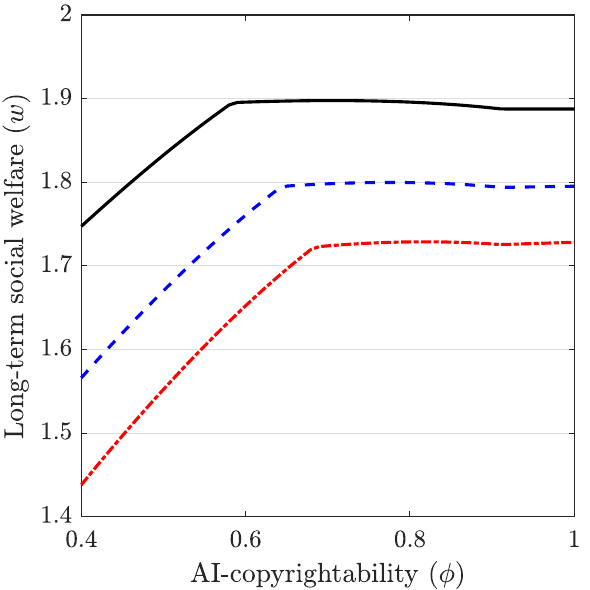}
			\label{fig:abundant_interaction_w}
		} 
	\end{center}
	\vspace{2mm}
	{\footnotesize \noindent \underline{\textbf{Parameters}}: $c_H = 0.3$, $c_A = 0.05$, $\lambda = 0.6$, $\beta =0.6$, $M = 2$, $k = 5$, $X_1 = 0.4$.
	} 
\end{figure}

To jointly assess the impact of the fair-use standard and AI-copyrightability, we combine the above findings to obtain the following result.
\begin{corollary}\label{coro:abundant_interaction}
In the data abundant regime, if $X_1$ is sufficiently large, the copyright policy that maximizes long-term creator income ($u$) and social welfare ($w$) is generous fair use ($f = 0$) and full AI-copyrightability ($\phi = 1$).
\end{corollary}
Figure \ref{fig:abundant_interaction} further illustrates the joint effect of fair use standard and AI-copyrightability. Across the three principal policy goals -- AI development, (long-term) consumer surplus ($s$), and (long-term) social welfare ($w$), generous fair use consistently outperforms strict fair use cases for all levels of AI-copyrightability. However, However, the performance gap between different fair use parameters ($f = 0$ versus $0.1$ and $0.2$) narrows as $\phi$ increases, indicating that strong AI-copyrightability can partially substitute for generous fair use.

\subsubsection{Impact of Policy Objectives and Market and Technological Factors.} 
\begin{figure}[htp]
	\begin{center}
		\caption{Optimal level of AI-copyrightability ($\phi^*$) under the data abundant regime}
		\label{fig:abundant_factors}

        \vspace{3mm}
		\subfloat[Current Status of AI ($X_1$)]
		{
			\includegraphics[width=0.3\textwidth]{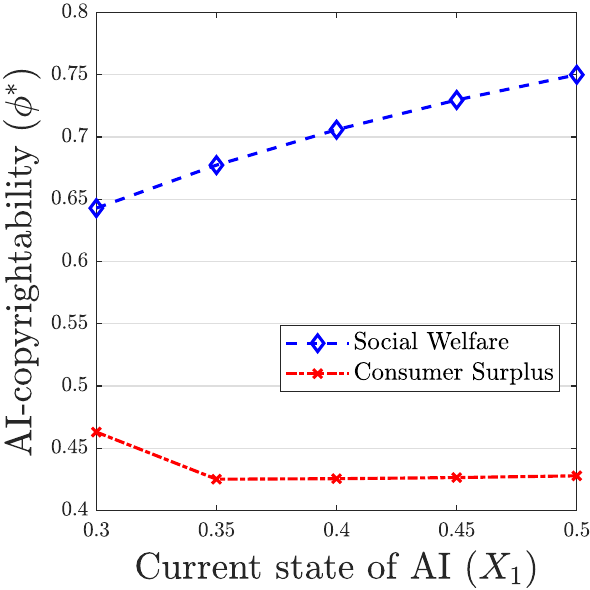}
			\label{fig:abundant_factors_X_1}
		} \quad \quad
        \subfloat[Market Growth ($M$)]
		{
			\includegraphics[width=0.3\textwidth]{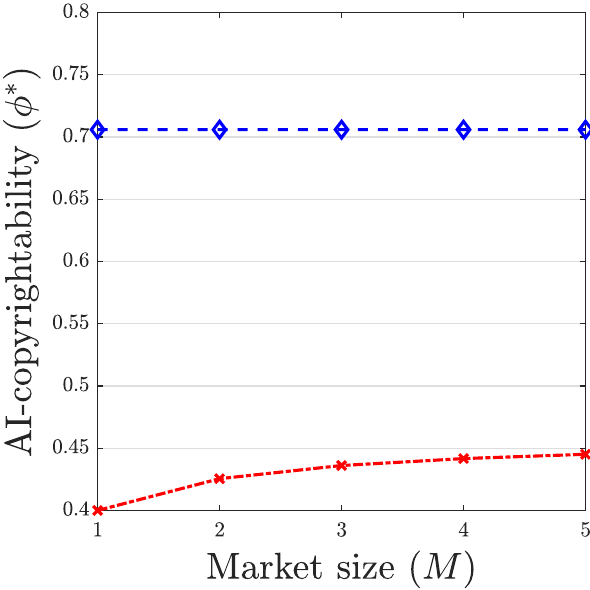}
			\label{fig:abundant_factors_M}
        }    \\
	\end{center}
	\vspace{2mm}
	{\footnotesize \noindent \underline{\textbf{Parameters}}: $c_H = 0.3$, $c_A = 0.05$, $\lambda = 0.6$, $\beta =0.6$, $f^* = 0$, $M = 2$ for (a), $X_1 = 0.4$ for (b).}
\end{figure}

Having examined the mechanisms behind the two policy levers, we now study how policy objectives and market and technological factors shape the optimal choice of policy parameters. Because generous fair use ($f = 0$) benefits all major stakeholders in the data abundant regime, we focus on the optimal choice of AI-copyrightability ($\phi^*$). Moreover, under generous fair use, the improved model quality satisfies $X_2^* = 1$ and is independent of $\phi$. We therefore denote attention to the two remaining objectives: maximizing social welfare and maximizing consumer surplus. For impacting factors, we consider the current capability of AI ($X_1$, Figure \ref{fig:abundant_factors_X_1}) and market growth ($M$, Figure \ref{fig:abundant_factors_M}).\footnote{In the parallel analysis for the data scarce case (\S \ref{sect:scarce_factor}), we also study the impact of training efficiency $k$. With abundant data, generous fair use is optimal regardless of $k$, rendering that parameter irrelevant here.
}

Several patterns are notable. First, policy objectives are the primary driver of $\phi^*$: as expected, the welfare-maximizing $\phi^*$ is significantly higher than the consumer-surplus-maximizing value. Second, the social-welfare maximizing $\phi^*$ increases with $X_1$, consistent with earlier findings (Figure \ref{fig:abundant_copyrightability_s_1} and Corollary~\ref{coro:abundant_interaction}) that stronger AI-copyrightability is more beneficial when the existing model quality is high. The impact of market growth ($M$) differs across policy objectives. Since $X_2^* = 1$, all creators produce AI-generated content in Period 2 under generous fair use; thus, the $\phi^*$ that is optimal for (long-term) social welfare depends only on Period-1 interactions and is independent of $M$. By contrast, for consumer surplus, an increase in $M$ raises the weight of Period-2 outcomes in determining $\phi^*$. Because high model quality tends to favor stronger AI-copyrightability, the consumer-surplus-maximizing $\phi^*$ increases as market prospects improve.

\subsection{Copyright Policy in the Data Scarce Regime}
As in the data abundant case, we start with the two copyright policies separately, followed by the resulting optimal policy choices under different technological, market, and regulatory conditions.

\subsubsection{Fair Use Standard.}
The impact of fair use standard in the data scarce regime is more complex. On the one hand, as in the data abundant regime, $f$ affects the cost to the AI company of acquire training data. On the other hand, it influences the supply of training data by shaping creators’ content-production decisions. Thus, unlike the data abundant regime -- where generous fair use always leads to a higher AI model quality-- the following proposition shows that this is not necessarily true when pre-existing data are scarce.
\begin{proposition}\label{prop:scarce_fair_use_generous}
In the data scarce regime, when $X_1$ is sufficiently large, the AI model quality in Period-2 under the strict fair use standard is (weakly) higher than that under generous fair use.
\end{proposition}
As shown, in the extreme case where the initial model quality $X_1$ is so high that creators do not produce human content under generous fair use, no new training data are available for model improvement. In contrast, under strict fair use, training compensation may induce some creators to produce human content, providing additional data for model training and yielding a higher model quality. This result highlights an important countervailing force: in the data scarce regime, generous fair use reduces the supply of training data. In this sense, its benefit of lowering acquisition costs can be offset by its adverse effect on data availability. However, the following proposition also reveals that overly strict fair use standard is not optimal either. 

\begin{proposition}\label{prop:scarce_fair_use_strict}
In the data scarce regime, when $f$ is sufficiently large, the AI firm does not improve model ($X_2 = X_1$). The Period-1 production policy and performance metrics are the same as those under generous fair use ($f = 0$). In Period-2, generous fair use leads to (i) (weakly) higher create income; and (ii) (weakly) higher social welfare when $X_1$ is sufficiently high. 
\end{proposition} 
The logic behind is intuitive: the training income received by human creators ($\frac{Q^C}{Q_{1H}} f$) depends on the fair use standard through two channels. Directly, holding demand for training data ($Q^C$) fixed, a higher $f$ raises training income. Indirectly, however, a higher $f$ increases the AI company’s cost of acquiring training data, which reduces its demand $Q^C$. When $f$ becomes prohibitively large, the AI company finds model improvement too costly and sets $Q^C = 0$. In that case, $f$ becomes irrelevant to creators, and the outcome reverts to the generous fair use benchmark.

\begin{figure}[htp]
	\begin{center}
		\caption{Impact of the fair use standard under the data scarce regime}
		\label{fig:scarce_fair_use}

        \vspace{3mm}
		\subfloat[AI development ($X_2$)]
		{
			\includegraphics[width=0.3\textwidth]{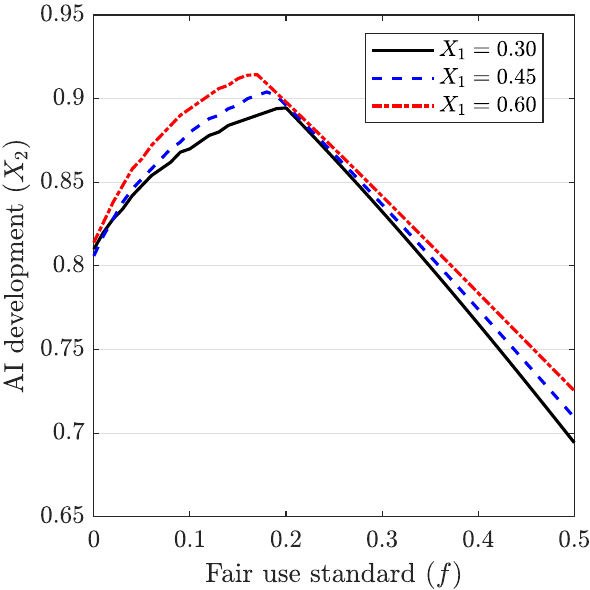}
			\label{fig:scarce_fair_use_X_2}
		} \quad 
		\subfloat[Period-2 social welfare ($w_2$)]
		{
			\includegraphics[width=0.3\textwidth]{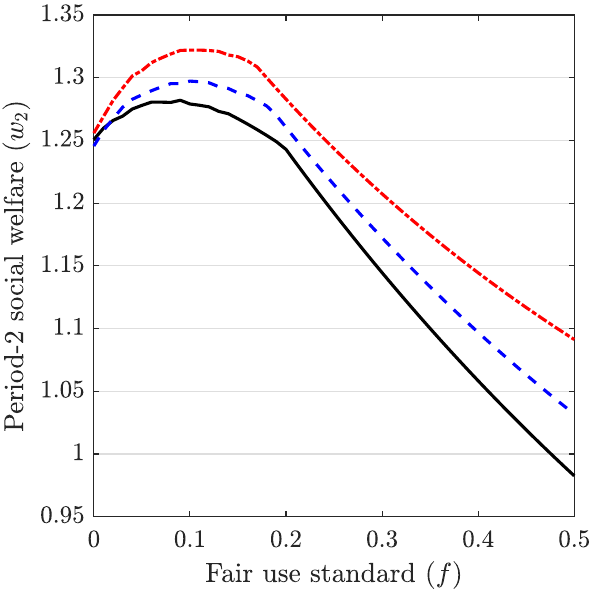}
			\label{fig:scarce_fair_use_w_2}
		} \quad 
        \subfloat[Period-1 social welfare ($w_1$)]
		{
			\includegraphics[width=0.3\textwidth]{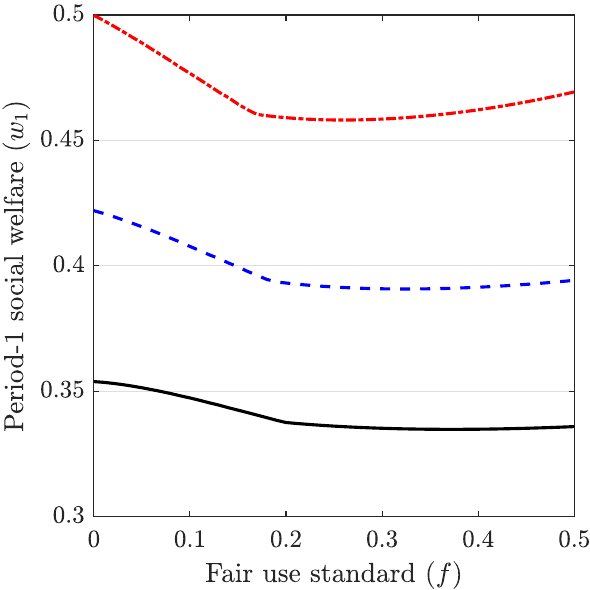}
			\label{fig:scarce_fair_use_w_1}
        }\\
        \vspace{1mm}
		\subfloat[Long-term social welfare ($w$)]
		{
			\includegraphics[width=0.3\textwidth]{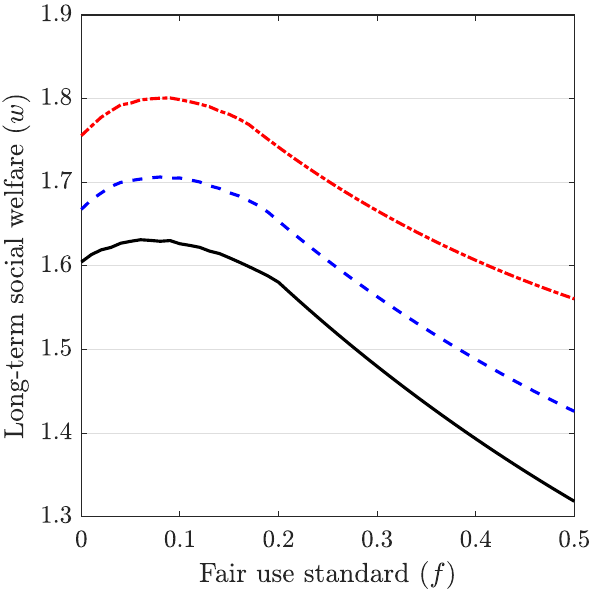}
			\label{fig:scarce_fair_use_w}
		} \quad
        \subfloat[Long-term consumer surplus ($s$)]
		{
			\includegraphics[width=0.3\textwidth]{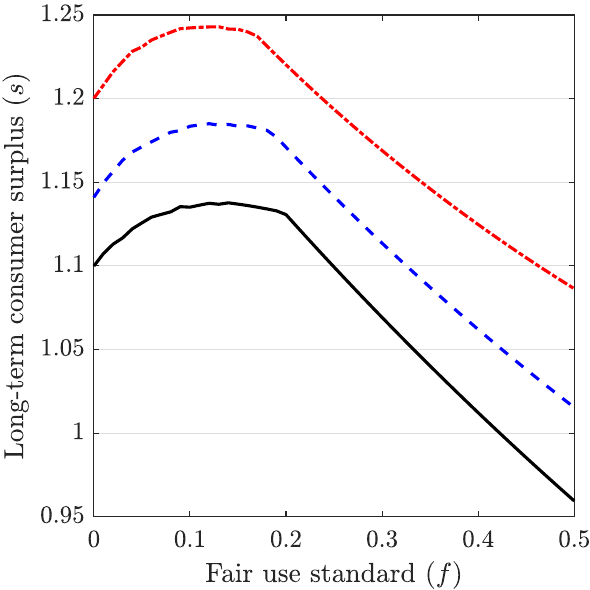}
			\label{fig:scarce_fair_use_s}
		}
	\end{center}
	\vspace{2mm}
	{\footnotesize \noindent \underline{\textbf{Parameters}}: $c_H = 0.3$, $c_A = 0.05$, $\lambda = 0.6$, $\beta =0.6$, $M = 2$, $k = 5$, $\phi = 0.6$. 
	} 
\end{figure}

Combining the two extreme scenarios in Propositions \ref{prop:scarce_fair_use_generous} and \ref{prop:scarce_fair_use_strict}, we expect a moderate fair use standard to be most beneficial for model development ($X_2$). The intuition is confirmed numerically in Figure~\ref{fig:scarce_fair_use_X_2}. In addition, the effect of the fair-use standard on Period-2 social welfare (Figure \ref{fig:scarce_fair_use_w_2} closely mirrors its impact on AI model development. As illustrated in Figure \ref{fig:scarce_fair_use_w_1}, fair use standard also affects Period-1 outcomes -- a phenomenon unique in the data scarce regime. Intuitively, stricter fair use rules encourage more human content creation than the consumption market would demand, potentially distorting short-term social welfare. However, as $f$ further increases, as the demand of training data from the AI firm drops, the effective income from training ($\frac{Q^C}{Q_{1H}} f$) also decreases, reversing this distortion. Taken together, a moderate fair use standard maximizes long-term social welfare (Figure \ref{fig:scarce_fair_use_w}) and consumer surplus (Figure \ref{fig:scarce_fair_use_s}), and this pattern is consistent across different initial model quality ($X_1$).

\subsubsection{AI-Copyrightability.}\label{sect:scarce_copyrightability} Similar to fair use standard, AI-copyrightability also influences model development by altering content creators' incentive to produce human content that can be used for model training. 
\begin{proposition}\label{prop:scarce_copyrightability}
In the data scarce regime, under generous fair use ($f=0$), stronger AI-copyrightability leads to more AI content and higher creator incomes in Period-1, but result in lower model quality in Period-2.
\end{proposition}
This proposition highlights both a similarity and a critical difference in the role of AI-copyrightability across the data abundant and data scarce regimes. As in the data abundant case, stronger AI-copyrightability directly encourages AI creation. However, in the data scarce regime, it simultaneously restricts the supply of training data, thereby undermining model development in the long run.

\begin{figure}[htp]
	\begin{center}
		\caption{Impact of AI-copyrightability under the data scarce regime}
		\label{fig:scarce_copyrightability}

        \vspace{3mm}
		\subfloat[AI Development ($X_2$)]
		{
			\includegraphics[width=0.3\textwidth]{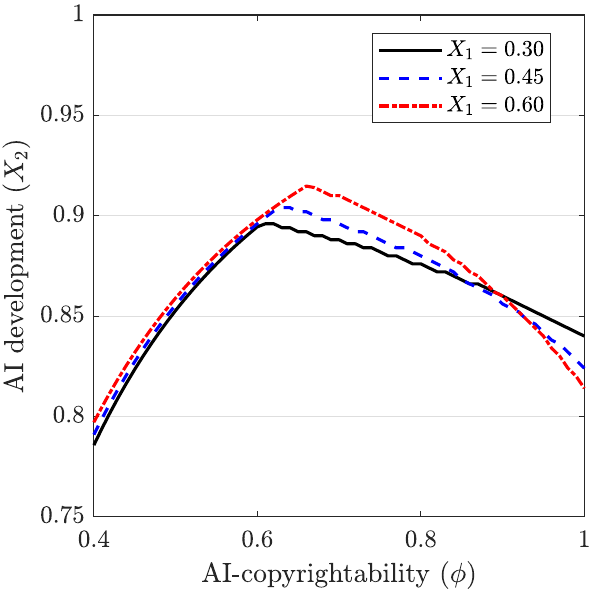}
			\label{fig:scarce_copyrightability_X_2}
		} \quad
		\subfloat[Period-2 Consumer Surplus ($s_2$)]
		{
			\includegraphics[width=0.3\textwidth]{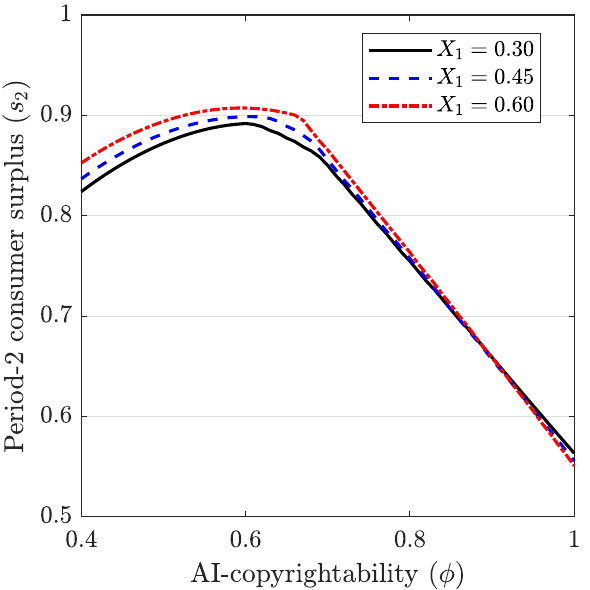}
			\label{fig:scarce_copyrightability_s_2}
		}\quad
        \subfloat[Period-2 Social Welfare ($w_2$)]
        {
			\includegraphics[width=0.3\textwidth]{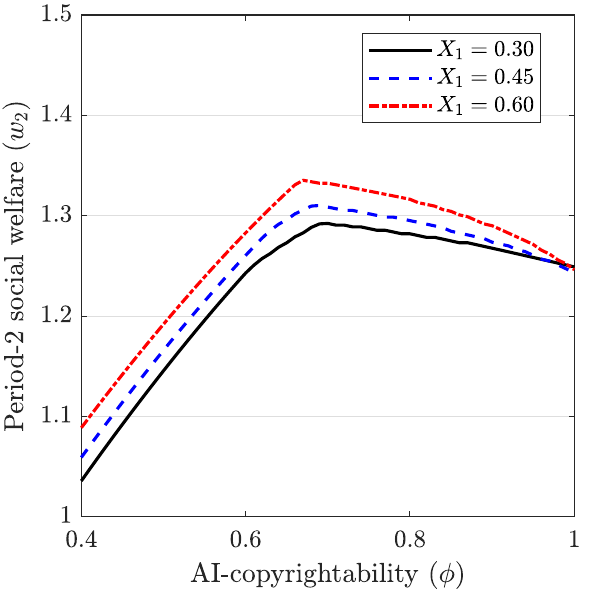}
			\label{fig:scarce_copyrightability_w_2}
		} 
	\end{center}
	\vspace{2mm}
	{\footnotesize \noindent \underline{\textbf{Parameters}}: $c_H = 0.3$, $c_A = 0.05$, $\lambda = 0.6$, $\beta =0.6$, $M = 2$, $k = 5$, $f = 0.2$.
	} 
\end{figure}

Under strict fair use ($f > 0$), AI-copyrightability affects model development through a different channel. Intuitively, when a higher $\phi$ increases the marginal benefit of improving model quality, the AI company demands more training data ($Q^C$). This, in turn, raises the expected training income for human creators. The resulting indirect effect relaxes the data-availability constraint and partially offsets the negative impact of AI copyrightability on model development. Taken together, AI-copyrightability has a more nuanced effect on model development in the data scarce regime than in the data abundant regime, reflecting the interplay between demand-side incentives and supply-side constraints. As illustrated in Figure \ref{fig:scarce_copyrightability_X_2}, the supply constraint becomes increasingly binding as $\phi$ increases, eventually dominating the demand-side boost. Consequently, $X_2$ first increases, and then decreases in $\phi$, with a steeper decline when $X_1$ is large. By comparing the metrics in Figure \ref{fig:scarce_copyrightability} to their counterparts in the data abundant case (Figure \ref{fig:abundant_copyrightability_2}), we observe similar patterns. 

\subsubsection{Interaction of Fair Use Standard and AI-Copyrightability.} 
\begin{figure}[htp]
	\begin{center}
		\caption{Interaction of fair use standard and AI-copyrightability under the data scarce regime}
		\label{fig:scarce_interaction}

        \vspace{3mm}
		\subfloat[AI development ($X_2$)]
		{
			\includegraphics[width=0.3\textwidth]{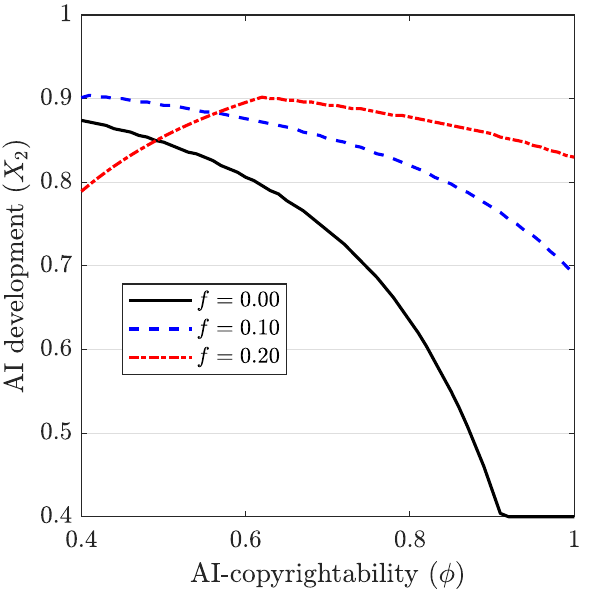}
			\label{fig:scarce_interaction_X_2}
		} \quad 
        \subfloat[Consumer surplus ($s$)]
		{
			\includegraphics[width=0.3\textwidth]{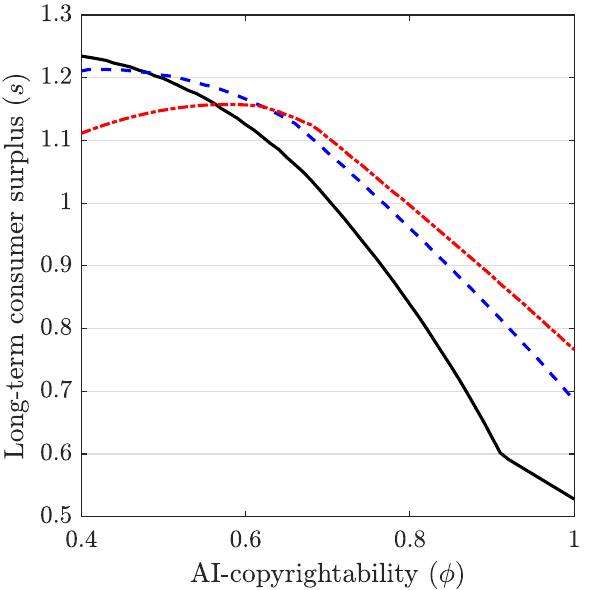}
			\label{fig:scarce_interaction_s}
		}\quad
		\subfloat[Social welfare ($w$)]
		{
			\includegraphics[width=0.3\textwidth]{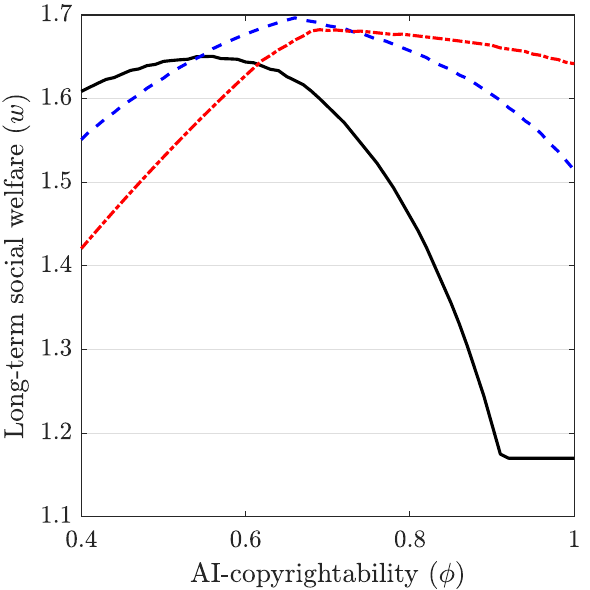}
			\label{fig:scarce_interaction_w}
		}
	\end{center}
	\vspace{2mm}
	{\footnotesize \noindent \underline{\textbf{Parameters}}: $c_H = 0.3$, $c_A = 0.05$, $\lambda = 0.6$, $\beta =0.6$, $M = 2$, $k = 5$, $X_1 = 0.4$.
	} 
\end{figure}

Regarding their interaction, recall that in the data abundant regime, these two policy function as weakly substitutes, yet optimal decisions could generally be made separately. When training data becomes scarce, however, the discussion on the effect of AI-copyrightability (\S\ref{sect:scarce_copyrightability}) suggests that this separation may break down. As shown in Figure \ref{fig:scarce_interaction_X_2}, when AI-copyrightability is weak, which nudges content creators toward generating human content and thus eases the supply constraint on training data, generous fair use is advantageous because it lowers model training costs. However, as AI-generated content receives stronger copyright protection, stricter fair use standard, which counteract the discouraging impact of AI-copyrightability on content creation, becomes more beneficial. The same holds for consumer surplus and social welfare (Figures \ref{fig:scarce_interaction_s} and \ref{fig:scarce_interaction_w}). 

Our findings suggest that, through the inter-temporal connection between the two periods, both policies have demand- and supply-side implications, and thus need to be considered jointly when training data becomes scarce. This interaction becomes particularly relevant when regulators face constraints in setting policies. For instance, leading US scholars have argued that AI model training likely constitutes fair use under existing US copyright laws in most circumstances \citep{henderson2023foundation}, which corresponds to $f = 0$ in our model. In that case, it might not make sense to grant AI-generated content too much copyright protection in the United States. Indeed, when the policy goal is AI development, our result reveals that under generous fair use, the level of AI-copyrightability should be set at the lower end ($\phi = \underline{\phi}$). That said, by comparing outcomes when a generous fair use constrain is imposed ($f = 0$) to the unconstrained cases ($f > 0$), we observe significant losses in AI development and social welfare caused by this constraint.

\subsubsection{Impact of Policy Objectives and Market and Technological Factors.}\label{sect:scarce_factor} 
\begin{figure}[htp]
	\begin{center}
		\caption{Optimal levels of policy parameters $(f^*, \phi^*)$ under the data scarce regime}
		\label{fig:scarce_factor}

        \vspace{3mm}
		\subfloat[$f^*$ on $X_1$]
		{
			\includegraphics[width=0.3\textwidth]{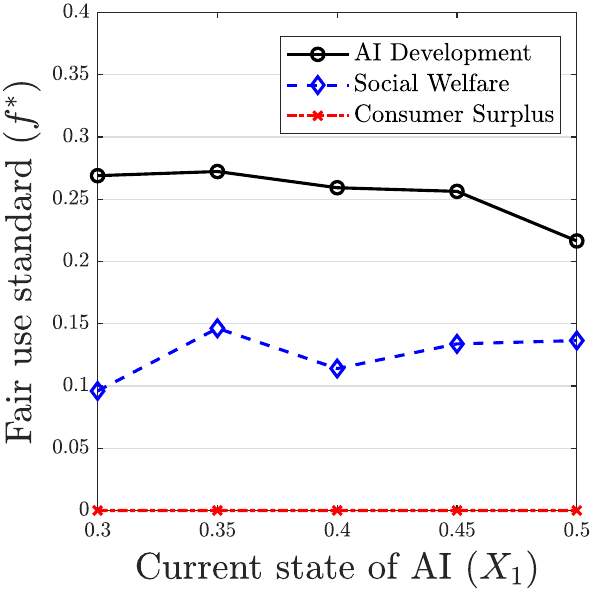}
			\label{fig:scarce_factor_f_X_1}
		} \quad
        \subfloat[$f^*$ on $M$]
		{
			\includegraphics[width=0.3\textwidth]{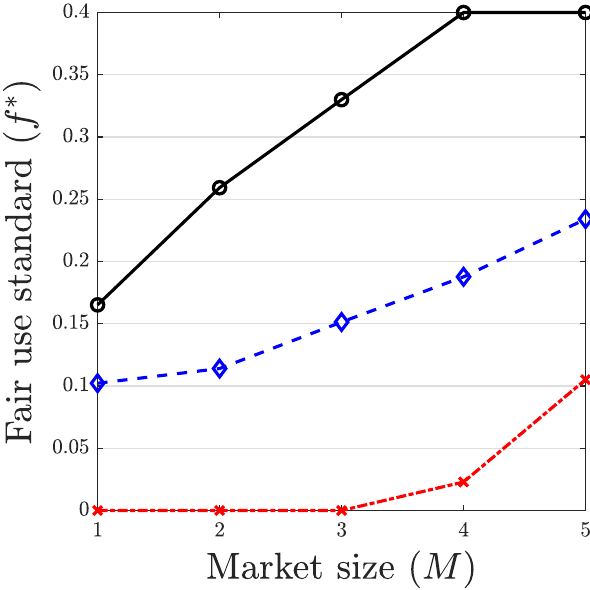}
			\label{fig:scarce_factor_f_M}
        }    
            \quad
		\subfloat[$f^*$ on $k$]
		{
			\includegraphics[width=0.3\textwidth]{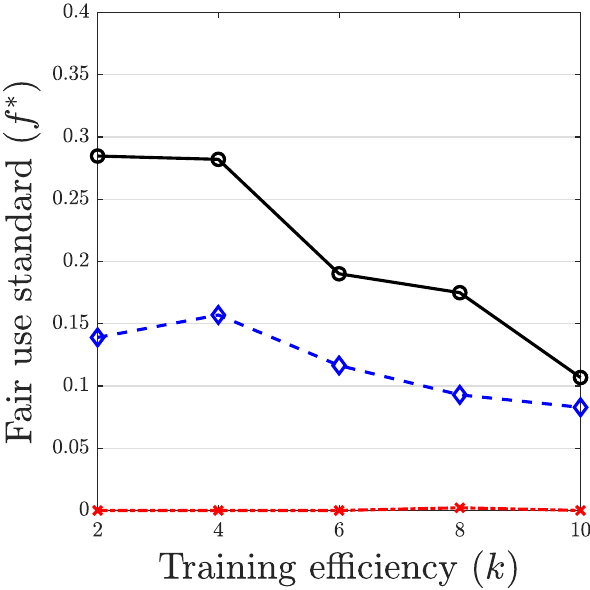}
			\label{fig:scarce_factor_f_k}
		}  \\
              \vspace{1mm}
		\subfloat[$\phi^*$ on $X_1$]
		{
			\includegraphics[width=0.3\textwidth]{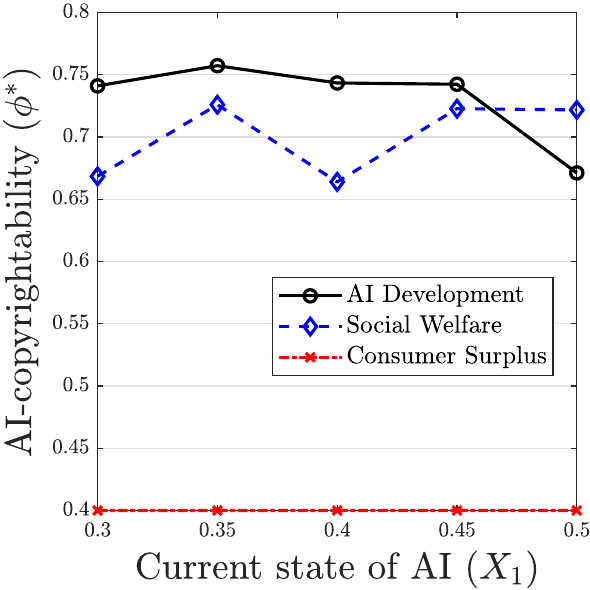}
			\label{fig:scarce_factor_phi_X_1}
		} \quad
        \subfloat[$\phi^*$ on $M$]
		{
			\includegraphics[width=0.3\textwidth]{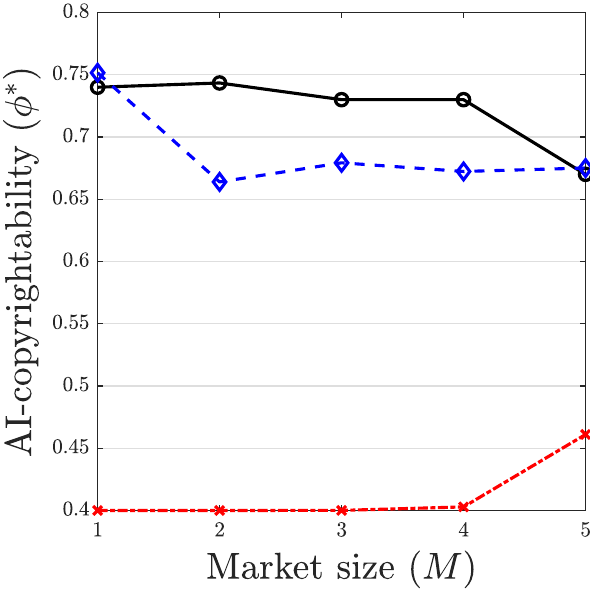}
			\label{fig:scarce_factor_phi_M}
        }    
            \quad
		\subfloat[$\phi^*$ on $k$]
		{
			\includegraphics[width=0.3\textwidth]{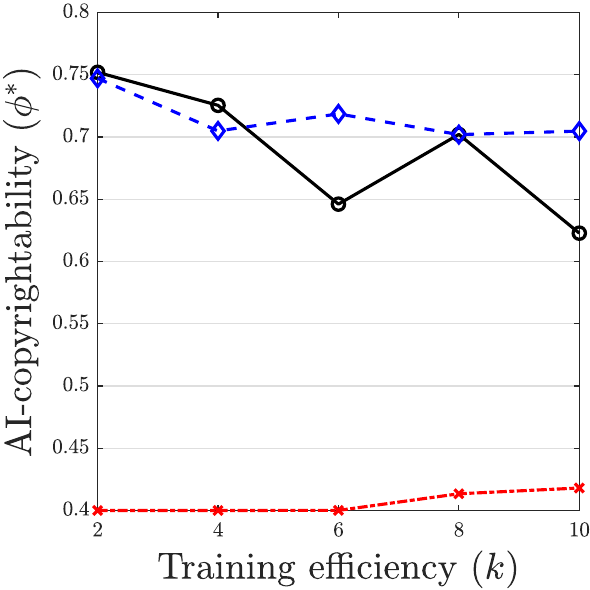}
			\label{fig:scarce_factor_phi_k}
		} 
	\end{center}
	\vspace{2mm}
	{\footnotesize \noindent \underline{\textbf{Parameters}}: $c_H = 0.3$, $c_A = 0.05$, $\lambda = 0.6$, $\beta =0.6$; $X_1 = 0.4$ for (b), (c), (e), (f); $M = 2$ for (a), (c), (d), (f); and $k = 5$ for (a), (b), (d), and (e).
	} 
\end{figure}

Different from the data abundant case, where generous fair use ($f = 0$) is optimal regardless of other factors, in the data scarce regime, both policy parameters are affected by policy objectives and other factors. Figure \ref{fig:scarce_factor} we present these impacts. In addition to $X_1$ (Figures \ref{fig:scarce_factor_f_X_1} and \ref{fig:scarce_factor_phi_X_1}) and $M$ (Figures \ref{fig:scarce_factor_f_M} and \ref{fig:scarce_factor_phi_M}), we examine the impact of training efficiency $k$, which captures how much data are required for model improvement (Figures \ref{fig:scarce_factor_f_k} and \ref{fig:scarce_factor_phi_k}). 

Several pattern emerge. First, as in the data abundant case, policy objectives play a primary role in determining both policy levers. Regarding AI-copyrightability, a regulator prioritizing consumer surplus consistently chooses the lowest value of $\phi$, while a regulator prioritizing social welfare or AI development prefers a substantially higher level. For fair use standard, the preferred level of $f$ is highest for a policymaker focused on AI development, and lowest for one focused on consumer surplus. The rationale is as follows: in our model, the main benefit of increasing $f$ is to induce additional human-generated content for model training, which directly enhances AI development. However, imposing $f > 0$ also encourages creators to consider factors beyond consumption value when choosing a production mode, potentially harming consumers. We also observe that while the gap in optimal $\phi$ between AI-development and welfare objectives is modest, the gap in optimal $f$ is much larger. This reflects the different underlying mechanisms behind the two policy levers: compared to AI-copyrightability -- which introduces two countervailing effects and therefore has a more nuanced impact -- the fair-use standard directly influences training-data demand with supply, making it a more effective tool.

These insights are further reinforced by the responsiveness of each policy lever to technological and market changes. While optimal $\phi$ remains relatively stable as $X_1$, $M$, and $k$ vary, the optimal fair use standard responds noticeably. For example, as the market grows -- an increasingly likely scenario as AI diffuses across industries—training-data demand rises, pushing the optimal $f$ upward. Conversely, if training becomes significantly more efficient (higher $k$), the demand for training data decreases, allowing $f$ to be set much lower. 

A comparison of the data-scarce and data-abundant regimes reinforces the broader conclusion that AI-copyrightability is largely driven by policy objectives and only weakly influenced by data availability. Importantly, unless policymakers focus exclusively on consumer surplus, some level of AI-copyrightability improves both social welfare and AI development. By contrast, the optimal fair use standard is more sensitive to changes in data availability, as well as shifts in other technological and market conditions.

\section{Concluding Remarks}
In this paper, we develop an equilibrium model to analyze how two key copyright policies related to generative AI -- the fair-use standard and AI-copyrightability -- shape AI development and stakeholder welfare. The insights carry important implications for policymakers navigating the evolving landscape of generative AI, at two levels. At a broader level, by capturing the dynamic nature of AI development and the unique supply-chain relationship between AI companies and content creators, we offer a framework that reveals the nuanced tradeoffs that copyright policies must confront. These tradeoffs imply that there is no one-size-fits-all doctrine for addressing these AI-related copyright issues. Instead, a flexible and agile governance approach -- attuned to unique institutional contexts -- is preferable.\footnote{In practice, regulatory adaptability emerges through several mechanisms. In the United States, where most intellectual property and AI disputes are advancing through private litigation, courts retain significant discretion in applying statutory and doctrinal standards to fact-specific disputes. Judicial discretion also extends to remedies: courts may vary damages, choose between statutory and actual compensation, or tailor injunctive relief based on case-specific considerations.} Moreover, by uncovering how fair use and AI-copyrightability interact, the paper shows that regulators, particularly those constrained by existing jurisprudence, should consider these two policy levers jointly rather than in isolation. 

At a more operational level, our findings provide guidance on how to tailor copyright regulation to technological conditions, market environments, and policy goals. For example, training-data availability plays a central role. When abundant training data are available, generous fair use is optimal, and AI-copyrightability can be determined separately based on policy objectives. As training data become scarce, however, the two levers must be jointly optimized. It is important to recognize that in practice, these regimes are not static. A model currently operating in the data abundant regime may transition the data-scarce regime as AI development progresses and training data are depleted. Accordingly, the results from the data-scarce regime are likely to become increasingly relevant as generative AI matures. Policy priorities matter as well. For example, China and the United States, the two AI superpowers, may lean towards prioritizing AI growth, whereas the European Union is more inclined to focus on the welfare of their consumers. Our findings suggests that these difference in policy objectives are likely to drive variation in these two regulatory levers. Finally, because its mechanism is direct, we expect the fair use standard to serve as a more agile tool for regulators as technological environments and market conditions evolve.

Beyond their policy implications, our findings highlight the importance for business leaders to grasp the intricate impact of AI-related regulatory decisions on their operational landscape. Policies designed to promote AI development may inadvertently diminish the supply of high-quality human-generated content, thereby affecting long-term innovation and data availability. Further, companies should anticipate regulatory divergence across jurisdictions and prepare for heterogeneous compliance environments. Finally, because industry stakeholders can shape the policy debate through lobbying and public engagement \citep{mackrael2023ai, tracy2023ai, zhang2024high}, our results suggest that AI companies have incentives to advocate for regulatory frameworks that preserve adaptive, responsive governance over static or rigid rules.

As one of the first analytical investigations of copyright regulation in the setting of generative AI, this paper opens several avenues for future research. First, in order to focus on the interaction between AI companies and content creators -- who are more directly influenced by copyright policies, we only model content creators as users of the AI model and do not consider end consumers directly engaging with AI service. A richer model incorporating consumers' interaction with AI services could yield additional insights. Second, our paper assumes that AI-generated content can be distinguished from human-created one. In practice, identification may be imperfect for technical reasons. Understanding strategic interactions under imperfect detection would be an important extension. Third, because our focus is on copyright regulation, the paper treats human-generated content as the sole source of model improvement. In practice, however, feedback data generated during AI usage -- such as user interactions and evaluations -- can also play an important role in training. Designing incentives that jointly optimize original content and feedback data would be valuable. Finally, our analysis focuses on a single regulatory body. Extending the model to include multiple regulators could shed light on topics such as innovation arbitrage \citep{samuelson2023generative}, forum shopping, and cross-border policy spillovers.

\bibliographystyle{informs2014} 
\bibliography{ref_alexyang,ref_GAI}

\newpage
\clearpage

\begin{APPENDICES}

\section{Definition of $(\underline{x}^S, \overline{x}^S)$ in Proposition \ref{prop:first_period_strict} and Brief Interpretation.}\label{appx:first_period_thresholds}

Define
\begin{align}
x_{HO}^F := \frac{c_H}{\beta + f};  \quad x_{HA}^F := \min\left\{1, \frac{c_H - c_A + \lambda \phi \beta X_1}{[1 - (1-\lambda)\phi]\beta + f}\right\}.
\end{align}
$x_{HO}^F$ (resp. $x_{HA}^F$) represent the threshold skill level above which a creator will prefer human creation over no creation (resp. AI creation) under full compensation. Note that $x_{HA}^F \ge x_{HO}^F$ if and only if $X_1 \ge X_A^F$, where
\begin{align}
X_A^F := \frac{c_A - (1-\lambda)\phi \frac{\beta}{\beta +f}c_H}{\lambda \phi \beta}.
\end{align}
Based on $x_{HO}^F$ and $x_{HA}^F$, define
\begin{align}\label{eq:X_F}
X^F &:= q^{-1}\left(Q_0 + \frac{1 - [\max(x_{HO}^F, x_{HA}^F)]^2}{2}; X_1\right), 
\end{align}
where $q^{-1}(\cdot, X_1)$ denotes the inverse of $q(\cdot; X_1)$ with respect to the first argument. $X^F$ represents the improved model quality if all human content created when the creators are fully compensated -- that is, they make their decisions based on the human-creation utility function $u_H^F = (\beta + f) x - c_H$ -- are used for model training. Next, let 
\begin{align}\label{eq:Omega_C}
\Omega^C = \{X \in [X_1, X^F] \; | \; \max_{p\ge 0} \pi_2(X,p) \ge 0\},
\end{align}
which represents the feasible region for improved model qualities {\em constrained} by the amount of data created under the above (hypothetical) human content creation policy. Accordingly, let $X^C$ represent the optimal solution under the feasible region, that is,
\begin{align}\label{eq:X_C}
X^C = \arg\max_{X \in \Omega^C} \phi \beta \lambda X - p_2^X(X);    
\end{align}
If $X^C \in (X_1, X^F)$, let $Q^C = q(X^C; X_1)$, which represents the amount of human content needed to improve the model quality to $X^C$. Based on $Q^C$, define $x_{HO}^{P}$ through the following implicit function:
\begin{align}\label{eq:x_HO_P}
\left[\beta + \frac{2 Q^C}{2 Q_0 + 1-\left(x_{HO}^{P}\right)^2}f \right] x_{HO}^{P} - c_H = 0.
\end{align}
Further, define $X_H^P$ as
\begin{align}
X_H^P := \frac{\left[\beta - (1-\lambda)\phi \beta + \frac{Q^C}{Q_0} f \right] - (c_H - c_A)}{\lambda \phi \beta}. 
\end{align} 
Based on $Q^C$ and $X_H^P$, we define threshold $x_{HA}^P$ as follows:
\begin{enumerate}[(i)]
\item when $X_1 \ge X_H^P$, $x_{HA}^P = 1$;
\item when $X_1 < X_H^P$, $x_{HA}^P$ is defined through the following implicit function:
\begin{align}\label{eq:x_HA_P}
\left[\beta - (1-\lambda)\phi \beta + \frac{2Q^C}{2 Q_0 + 1-\left(x_{HA}^{P}\right)^2} f \right] x_{HA}^{P} - \lambda \phi \beta X_1 - (c_H - c_A) = 0.
\end{align}
\end{enumerate}
$x_{HO}^P$ ($x_{HA}^P$) represents the threshold skill level above which a creator will prefer human creation over no creation (AI creation) under partial compensation. $x_{HA}^P \ge x_{HO}^P$ if and only if $X_1 \ge X_A^P$, where
\begin{align}
X_A^P 
:= \frac{c_A - (1-\lambda)\phi \beta x_{HO}^{P}}{\lambda \phi \beta} \label{eq:X_A_P}.
\end{align}
Under these definitions and $x_{AO}$ defined in Eq. \eqref{eq:threshold_skill} ($x_{AO} = \frac{c_A - \lambda \phi \beta X_1}{(1-\lambda \phi \beta)}$ under $X_1$ and $p_1 = 0$), we have:
\begin{enumerate}
\item If $X^C = X_1$, the firm uses no data for model improvement, and human content creators receive no income from model training. $\underline{x}^S$ and $\overline{x}^S$ are the same as those defined in Proposition \ref{prop:first_period_abundant}.
\item If $X^C \in (X_1, X^F)$, the firm uses part of the human content created for model improvement, and human content creators receive partial training income ($ < f$). The thresholds that define content creation policy are:
\begin{enumerate}
    \item for $X_1 \le X_A^{P}$, $\underline{x}^S = \overline{x}^S =  x_{HO}^{P}$;
    \item for $X_1 > X_A^{P}$, $\underline{x}^S = \max(0, x_{AO})$ and $\overline{x}^S = x_{HA}^{P}$.
\end{enumerate}
\item If $X^C = X^F$, the AI company uses all data created for model improvement, and human content creators receive full training income ($f$). The thresholds follow:
\begin{enumerate}
    \item for $X_1 \le X_A^F$, $\underline{x}^S = \overline{x}^S = x_{HO}^F$;
    \item for $X_1 > X_A^F$, $\underline{x}^S = \max(0, x_{AO})$ and $\overline{x}^S = x_{HA}^F$.
\end{enumerate}
\end{enumerate}

\newpage
\section{Proofs}


\subsubsection*{Proof of Proposition \ref{prop:second_period_production}.} 
For a creator with skill level $x$, the (net) income of creating human content is 
\begin{align}
u_H = \beta x - c_H.   
\end{align}
Next, facing AI model with quality $X$, the utility of creating AI content is:
\begin{align}
u_A= \phi \beta [\lambda X + (1-\lambda)x] - c_A.    
\end{align} 
Finally, $u_N = 0$ for creators not producing anything. 

To prove the double-threshold policy, we want to show that for $x_1 < x_2$, if a creator with skill level $x_1$ chooses human creation, then the creator with skill level $x_2$ will also chooses human creation. To see that, note that the creator with $x_1$ chooses human creation if and only if 
\begin{align}
u_H(x_1) > \max(u_A(x_1), 0),
\end{align}
then we have $u_H(x_2) > u_H(x_1) > 0$, and 
\begin{align}
u_H(x_2) \ge u_H(x_1) + \beta(x_2 - x_1) \ge u_A(x_1) + (1-\lambda)\phi \beta(x_2 - x_1) = u_A(x_2).
\end{align}
Thus, we have $u_H(x_2) > \max(u_A(x_2), 0)$, that is, a creator with $x_2$ also chooses human creation. Symmetrically, we could show that for $x_1 < x_2$, if creator with $x_2$ chooses to produce no content, then creator with $x_1$ also choose not to produce anything. Combining these two results, we obtain the double-threshold policy. 

Next, we establish the structure of the thresholds $\underline{x}$ and $\overline{x}$ as characterized in Table \ref{table:second_period_threshold}. Here, as mentioned above, a creator chooses to produce human content if and only if $u_H \ge \max(u_A, u_N)$. That is, 
\begin{align}
\beta x - c_H \ge \max\left(\phi \beta [\lambda X + (1-\lambda)x] - c_A, \; 0\right).
\end{align}
Or equivalently,  
\begin{align}
x \ge \max\left(x_{HO}, \;\; x_{HA} \right), 
\end{align}
where
\begin{align}
x_{HO} = \frac{c_H}{\beta}; \quad x_{HA} = \frac{\lambda \phi \beta X + (c_H - c_A) }{[1 - (1-\lambda)\phi] \beta}.
\end{align}
Similarly, a creator chooses to generate AI-content if and only if
\begin{align}
x \in \left(x_{AO}, \;\; x_{HA}\right). 
\end{align}
where $x_{HA}$ is defined above, and
\begin{align}
x_{AO} = \frac{c_A + p - \lambda \phi \beta X}{(1-\lambda) \phi \beta}.
\end{align}
Finally, a creator will choose not to enter the market if and only if
\begin{align}
x \le \min \left(x_{HO}, \;\; x_{AO} \right).
\end{align}
Let
\begin{align}
X_A := \frac{c_A + p - (1-\lambda)\phi c_H}{\lambda \phi \beta},
\end{align}
we have:
\begin{enumerate}
\item when $X < X_A$, it could be verified that $x_{HA} \le x_{HO} \le x_{AO}$, and the corresponding policy is:
\begin{enumerate}
\item for $x \in [0, x_{HO})$, creators choose not to produce any content.
\item for $x \in (x_{HO}, 1]$, creators choose to produce human content. 
\end{enumerate}
Note that as $c_H \in (0, \beta)$, we have $x_{HO} \in (0, 1)$, so both regions are non-empty. 
\item when $X \ge X_A$, it could be verified that $x_{AO} \le x_{HO} \le x_{HA}$, and the corresponding content production policy is
\begin{enumerate}
\item for $x \in [0, x_{AO})$, creators choose not to produce any content.
\item for $x \in [x_{AO}, x_{HA}]$, creators choose to produce AI content
\item for $x \in (x_{HA}, 1]$, creators choose to produce human content.
\end{enumerate}
\end{enumerate}
To further refine the three regions in the second scenario ($X \ge X_A$), we could show that when 
\begin{align}
X \ge X_H := 1 - \frac{(c_H - c_A - p) - (1 - \phi)\beta}{\lambda \phi \beta}.
\end{align}
We have $x_{HA} > 1$, thus there is no human creation. Similarly, when
\begin{align}
X \ge X_N := \frac{c_A + p}{\lambda \phi \beta},
\end{align}
we have $x_{AO} < 0$, thus every creator produces some content. Further, by comparing $X_A$, $X_N$, and $X_H$, we have that $X_A < \min(X_N, X_H)$ and $X_H > X_N$ if and only if $c_H < [1 - (1-\lambda)\phi] \beta$.
Reorganizing the above scenarios lead to the thresholds and cases in Table \ref{table:second_period_threshold}. \hfill $\square$



\medskip

\subsubsection*{Proof of Corollary \ref{coro:second_period_performance}.}  
Under the above policy, the number of (high-quality) human content is
\begin{align}
Q_{2H} = M \int_{\overline{x}}^1 x dF(x) = \frac{M(1-\overline{x}^2)}{2}.
\end{align}
The amount of AI content is:
\begin{align}
Q_{2A} = M \int_{\underline{x}}^{\overline{x}}[\lambda X + (1-\lambda)x]dF(x) = M \left[\lambda X (\overline{x}-\underline{x}) + \frac{1-\lambda}{2} \left(\overline{x}^2 - \underline{x}^2\right) \right].
\end{align}
The AI company's revenue is:
\begin{align}
r_2 = M \int_{\underline{x}}^{\overline{x}} p_2 dF(x) = M p_2 (\overline{x} - \underline{x}).
\end{align}
The human content creators' income is:
\begin{align}
u_{2H} = \beta Q_H - M c_H (1 - \overline{x}) = M\left[\frac{\beta}{2} \left(1-\overline{x}^2\right) - c_H (1-\overline{x})\right].
\end{align}
The total AI-content creators' profit is:
\begin{align}
u_{2A} = \phi \beta Q_A - M c_A (\overline{x} - \underline{x}) = M \left[\phi \beta \left(\lambda X (\overline{x}-\underline{x}) + \frac{1-\lambda}{2} \left(\overline{x}^2 - \underline{x}^2\right) \right)- c_A (\overline{x}-\underline{x}) \right].
\end{align}
And the content creators' aggregated income total creators' profit is:
\begin{align}
u_2 = u_{2H} + u_{2A} = \beta(Q_{2H} + \phi Q_{2A}) - M \left[c_H (1-\overline{x}) + c_A (\overline{x} - \underline{x})\right].
\end{align}
The total consumer surplus is:
\begin{align}
s_2 = (1-\beta) Q_{2H} + (1-\phi\beta) Q_{2A}.
\end{align}
For social welfare, we have
\begin{align}
w_2 = u_2 + s_2 + r_2 = Q_{2H} + Q_{2A} - M \left[c_H (1-\overline{x}) + c_A (\overline{x} - \underline{x}) \right]
\end{align}

Next, we consider results related to monotonicity. First, note that as $\overline{x}$ increases concavely in $X_2$ and $\underline{x}$ decreases convexly in $X_2$, we have that $Q_{2A}$ -- which increases in $\overline{x}$ and decreases in $\underline{x}$ -- increases in $X_2$. Similarly, $r_2$, which increases linearly in $\overline{x} - \underline{x}$, increases concavely in $X_2$. 

For the monotonicity of $u_2$, note that we could re-write a single creator's optimal content production policy as $\max(u_H, u_A, 0)$, which (weakly) increases in $X_2$ through $u_A$. By writing $u_2$ as
\begin{align}
u_2 = M \int_0^1 \max\left( u_H, u_A(X_2), 0\right) dx,
\end{align}
we can see that $u_2$ also increases in $X_2$. 

Finally, to prove that $r_2$ is concave on $p_2$, we can first verify when $X_2 \ge X_A(0)$, $r_2$ is positive at $p_2 \in (0, \bar{p}(X_2))$ and equals to $zero$ at $p_2 = 0$ or $p_2 = \bar{p}(X_2)$. Within this interval, $r_2$ is a continuous piece-wise function of $p_2$, where the specific expressions depends on the magnitude of different parameters. For example, assume $\phi \le \frac{\beta - c_H}{\beta(1-\lambda)}$ (Case I in Table \ref{table:second_period_threshold}). Depending on the magnitude of $X_2$, $r_2$ can be written as the following piece-wise function of $p_2$ 
\begin{enumerate}
\item When $X_2 \in (X_A(0), X_N(0)]$, 
\begin{align}\label{eq:r_2_p_2_1}
r_2 = M p_2 [x_{HA}(X_2, p_2) - x_{AO}(X_2,p_2)]; 
\end{align}
\item When $X_2 \in (X_N(0), X_H(0)$, let $p_N = \lambda \phi \beta X - c_A$. We have:
\begin{align}\label{eq:r_2_p_2_2}
r_2 = 
\begin{cases}
M p_2 [x_{HA}(X_2, p_2)]; \quad &\mbox{if} \quad p_2 \in [0, p_N]; \\
M p_2 [x_{HA}(X_2, p_2) - x_{AO}(X_2,p_2)]; \quad &\mbox{if} \quad p_2 \in (p_N, \bar{p}(X_2)];
\end{cases}
\end{align}
\item When $X_2 > X_H(0)$, let $p_H = \lambda \phi \beta X - c_A + c_H - [(1-(1-\lambda)\phi]\beta$. We have:
\begin{align}\label{eq:r_2_p_2_3}
r_2 = 
\begin{cases}
M p_2; \quad &\mbox{if} \quad p_2 \in [0, p_H]; \\
M p_2 [x_{HA}(X_2, p_2)]; \quad &\mbox{if} \quad p_2 \in [p_H, p_N]; \\
M p_2 [x_{HA}(X_2, p_2) - x_{AO}(X_2,p_2)]; \quad &\mbox{if} \quad p_2 \in (p_N, \bar{p}(X_2)];
\end{cases}
\end{align}
\end{enumerate}
For the first scenario (Equation~\eqref{eq:r_2_p_2_1}), using quantities $x_{HA}$ and $x_{AO}$ defined in Table \ref{table:second_period_threshold}, we can verify that $r_2$ is quadratic on $p_2$ with negative second-order derivative. Thus, $r_2$ is concave in this scenario. For the other two cases (Equations \eqref{eq:r_2_p_2_2} and \eqref{eq:r_2_p_2_3}), we can similarly show that $r_2$ is quadratic on $p_2$ with negative second-order derivatives for both scenarios in Equation~\eqref{eq:r_2_p_2_2} and scenarios 2 and 3 in Equation ~\eqref{eq:r_2_p_2_2}. Finally, note that $r_2$ is linear on $p_2$ in the first scenario in Equation~\eqref{eq:r_2_p_2_3}. Thus, within each segment, $r_2$ is (weakly) concave on $p_2$. 

Next, we consider the concavity of $r_2$ at breakpoints ($p_H$ or $p_N$), where $r_2$ is continuous but not continuously differentiable. At $p_H$, we have:
\begin{align}
\frac{\partial r_2}{\partial p_2} \Big|_{p_2 = p_H^-} &= M; \\
\frac{\partial r_2}{\partial p_2} \Big|_{p_2 = p_H^+} &= M \left[ x_HA + p_2 \frac{\partial x_{HA}}{\partial p_2}\right] = M \left[1 - \frac{p_N}{[1-(1-\lambda)\phi] \beta} \right].
\end{align}
Therefore, 
\begin{align}
\frac{\partial r_2}{\partial p_2} \Big|_{p_2 = p_H^-} > \frac{\partial r_2}{\partial p_2} \Big|_{p_2 = p_H^+}.
\end{align}
Similarly, we can show that 
\begin{align}
\frac{\partial r_2}{\partial p_2} \Big|_{p_2 = p_N^-} > \frac{\partial r_2}{\partial p_2} \Big|_{p_2 = p_N^+}.
\end{align}
That is, at each breakpoint, $r_2$ is also concave on $p_2$. This completes the proof that $r_2$ is (weakly) concave on $p_2$ when $p_2 \in (0, \bar{p}(X_2))$ when $\phi \le \frac{\beta - c_H}{\beta(1-\lambda)}$. 

Similarly, when $\phi > \frac{\beta - c_H}{\beta(1-\lambda)}$ (Case II in Table \ref{table:second_period_threshold}), we can also show that $r_2$ is (weakly) concave on $p_2$ when $p_2 \in (0, \bar{p}(X_2))$. \hfill $\square$ 

\medskip

\subsubsection*{Proof of Proposition \ref{prop:model_improvement_generous}.}
According to Corollary \ref{coro:second_period_performance}, the AI company's revenue $r_2$ increases in $X_2$. On the other hand, the cost of acquiring training data is $0$. Thus, it is optimal for the AI company to utilize all available data, $Q_0 + Q_{1H}$, to improve the model, as characterized in Eq. \eqref{eq:X_2_G}. \hfill $\square$

\medskip

\subsubsection*{Proof of Proposition \ref{prop:model_improvement_strict_large_X}.}
Based on Corollary \ref{coro:second_period_performance}, $\pi_2$ is also concave on $p_2 \in [p_2^X(X_2), \; \overline{p}(X_2)]$ for any $X_2 \in \Omega_2(X_1)$. Further, by the definition of $p_2^X$, we have:
\begin{align}
\frac{\partial \pi_2(X_2, p_2^X)}{\partial p_2} > 0.
\end{align}
Similarly, based on Corollary \ref{coro:second_period_performance}, $r_2$ is concave on $X_2$ and $q(X_2; X_1)$ is convex on $X_2$. Thus $\pi_2 = r_2 - fq(X_2; X_1)$ is concave on $X_2$. Further, by the definition of $p_2^X$, we have: 
\begin{align}
\frac{\partial \pi_2(X_2, p_2^X)}{\partial X_2} < 0.
\end{align}
Therefore, for the monotonicity of $p_2^X$ on $X_2$, by the Implicit Function Theorem, we have:
\begin{align}
\frac{\partial p_2^X}{\partial X_2} = -\frac{\frac{\partial \pi_2(X_2, p_2^X)}{\partial X_2}}{\frac{\partial \pi_2(X_2, p_2^X)}{\partial p_2}} > 0.
\end{align}
For the convexity of $p_2^X$, by the Implicit Function Theorem, we have:
\begin{align}
\frac{\partial^2 p_2^X}{\partial X_2^2} = -\left(\frac{\partial \pi_2}{\partial p_2}\right)^{-1} \left[ \frac{\partial^2 \pi_2}{\partial X_2^2} + \frac{\partial^2 \pi_2}{\partial X_2 \partial p_2} \frac{\partial p_2^X}{\partial X_2} + \frac{\partial^2 \pi_2}{\partial X_2^2} \left( \frac{\partial p_2^X}{\partial X_2} \right)^2 \right].
\end{align}
Note that $\frac{\partial \pi_2}{\partial p_2} > 0$, and $\frac{\partial^2 \pi_2}{\partial X_2^2} = - q''(X_2; X_1)$. Therefore, when $q''(\cdot)$ is sufficiently large, 
\begin{align}
\frac{\partial^2 \pi_2}{\partial X_2^2} + \frac{\partial^2 \pi_2}{\partial X_2 \partial p_2} \frac{\partial p_2^X}{\partial X_2} + \frac{\partial^2 \pi_2}{\partial X_2^2} \left( \frac{\partial p_2^X}{\partial X_2} \right)^2 < 0, 
\end{align}
and hence $\frac{\partial^2 p_2^X}{\partial X_2^2} > 0$. 

Next, we prove the three scenarios for the optimal $(X_2, p_2)$ based on the magnitude of $\lambda \phi \beta$. Note that the objective is to maximize $\lambda \phi \beta X_2 - p_2^X(X_2)$, we have
\begin{align}
\frac{d}{d X_2} \left(\lambda \phi \beta X_2 - p_2^X(X_2)\right) = \lambda \phi \beta - \frac{\partial p_2^X}{\partial X_2}.   
\end{align}
Given that $\frac{\partial p_2^X}{\partial X_2}$ is increasing, we have if $\frac{\partial p_2^X}{\partial X_2} \ge \lambda \phi \beta$, the optimal solution is $X_2 = X_1$, and $p_2 = 0$. This corresponds to the first scenario.

On the other hand, if $\frac{\partial p_2^X}{\partial X_2}|_{X_2=\min(\overline{X}_2, X_2^D)} \le \lambda \phi \beta$, the optimal improved model quality should be as high as possible, that is, $X_2=\min(\overline{X}_2, X_2^D)$, the third scenario in the proposition.

Finally, when $\lambda \phi \beta \in \left(\frac{\partial p_2^X}{\partial X_2}|_{X_2=X_1}, \; \frac{\partial p_2^X}{\partial X_2}|_{X_2=\min(\overline{X}_2, X_2^D)} \right)$, the optimal improved model quality is the interior optimal. \hfill $\square$

\medskip

\subsubsection*{Proof of Proposition \ref{prop:first_period_abundant}.} The proof of the creation policy is the same as that in Proposition \ref{prop:second_period_production}. \hfill $\square$

\medskip

\subsubsection*{Proof of Proposition \ref{prop:first_period_strict}.}
The proof that the content production policy follows a double-threshold structure is similar to that in Proposition \ref{prop:second_period_production} and is omitted for brevity. In the following, we focus on characterizing the thresholds $\overline{x}^S$ and $\underline{x}^S$. To do so, we first define the following auxiliary utility functions for human content creation. Let $u_H^F(x)$ be a creator's utility of producing human content and receives the full training price $f$ (the superscript $F$ representing {\em full} compensation), that is,
\begin{align}
u_H^F(x) = (\beta + f) x - c_H;
\end{align}
Assuming the utility function of human creation is $u_H^F()$, by comparing that with $u_A$ and $u_O$ (note that the training income does not affect the utility function under AI-generation and no generation), the creators' content production policy is:
\begin{enumerate}
\item when $X_1 \le X_A^F$, we have: $x_{HO}^F \ge x_{HA}^F$. Thus, a creator produces human content when $x \ge x_{HO}^F$ and does not create any content for $x < x_{HO}^F$. 
\item when $X_1 > X_A^F$, we have $x_{HO}^F < x_{HA}^F$. Thus, a creator produces human content when $x \ge x_{HA}^F$, AI content when $x \in (x_{AO}, x_{HA}^F)$, and does not create any content for $x < x_{AO}$. 
\end{enumerate}
By combining these two scenarios, the amount of human content created in Period-1 is:
\begin{align}\label{eq:Q_1H_F}
Q_{1H}^F = \frac{1-[\max(x_{HO}^F, x_{HA}^F]^2}{2}.
\end{align}
and the highest model quality that can be achieved given $Q_{1H}^F$ and $Q_0$, which is also the highest model quality given $f$, is:
\begin{align}
X^F = q^{-1}(Q_0 + Q_{1H}^F; X_1),
\end{align}
as defined in Equation \eqref{eq:X_F}. Accordingly, $\Omega^C$ (Equation \eqref{eq:Omega_C}) is the feasible region for improved model qualities constrained by the maximum amount of human data created in Period-1 based on the copyright policy ($f$), and $X^C$ (Equation \eqref{eq:X_C}) is the optimal solution within this feasible region. Further, $Q^C =q(X^C; X_1)$ is the amount of data required to improve the model quality from $X_1$ to $X^C$. 

Based on these definitions, we characterize $\underline{x}^S$ and $\overline{x}^S$ by the following three scenarios. 

\noindent {\bf Scenario I: $X^C = X_1$.} In this scenario, we can verify that in anticipation that the that the AI company will not acquire any data for model training, creators will make their creation mode choice based on the human creation utility function $u^H = x - c_H$. As a result, the content creation policy follows the same as in the generous fair use case ($f = 0$), that is, $\overline{x}^S = \overline{x}(X_1, 0)$ and $\underline{x}^S = \underline{x}(X_1, 0)$. This corresponds to Case 1 in Appendix \ref{appx:first_period_thresholds}.

\noindent {\bf Scenario II: $X^C = X^F$.} In this scenario, we consider the following two cases:
\begin{enumerate}
\item When $X_1 \le X_A^F$, we can verify that $x_{HA}^F \le x_{HO}^F$. Therefore, 
\begin{align}
X^C = X^F = q^{-1}\left(Q_0 + \frac{1 - (x_{HO}^F)^2}{2}; X_1\right),
\end{align}
and thus, 
\begin{align}
Q^C = \frac{1 - (x_{HO}^F)^2}{2}.
\end{align}
We can then verify that $u_H^F(x) \ge \max(u_A(x), 0)$ if and only if $x \ge x_{HO}^F$. That is, under $u_H^F(x)$, only creators with $x \ge x_{HO}^F$ choose human creation. By Equation~\eqref{eq:Q_1H_F}, we have:
\begin{align}
Q_{1H}^F = \frac{1 - (x_{HO}^F)^2}{2} = Q^C.
\end{align}
This confirms that under the above human creation policy, the amount of human creator content will all be used in Period-2 model training, which leads to the human creation utility function $u_H^F(x)$. Therefore, $\overline{x}^S = x_{HO}^F$ is the equilibrium human creation threshold. Accordingly, we can show that $0 > \max(u_H^F(x), u_A(x))$ if and only if $x < x_{HO}^F$. Thus, $\underline{x}^S = x_{HO}^F$ is another equilibrium threshold. 
\item When $X_1 > X_A^F$, we can verify that $x_{HA}^F > x_{HO}^F$. Therefore, 
\begin{align}
X^C = X^F = q^{-1}\left(Q_0 + \frac{1 - (x_{HA}^F)^2}{2}; X_1\right),
\end{align}
and thus, 
\begin{align}
Q^C = \frac{1 - (x_{HA}^F)^2}{2}.
\end{align}
We can then verify that $u_H^F(x) \ge \max(u_A(x), 0)$ if and only if $x \ge x_{HA}^F$. That is, under $u_H^F(x)$, only creators with $x \ge x_{HA}^F$ choose human creation. By Equation~\eqref{eq:Q_1H_F}, we have:
\begin{align}
Q_{1H}^F = \frac{1 - (x_{HA}^F)^2}{2} = Q^C.
\end{align}
This confirms that under the above human creation policy, all human creator content will be used in Period-2 model training, which leads to the human creation utility function $u_H^F(x)$. Therefore, $\overline{x}^S = x_{HA}^F$ is the equilibrium human creation threshold. Accordingly, we can show that $0 > \max(u_H^F(x), u_A(x))$ if and only if $x < \max(0, x_{AO})$. Thus, $\underline{x}^S = \max(0, x_{AO})$ is another equilibrium threshold. 
\end{enumerate}
This corresponds to case 3 in the Appendix.

\noindent {\bf Scenario III: $X^C \in (X_1, X^F)$.} In this case, the thresholds -- $(\underline{x}^S, \overline{x}^S)$ -- are determined by the creators' rational expectation on the amount of data that the AI company will acquire for Period-2 model training ($Q^C$), as well as that on the aggregated human content created based on the double threshold creation policy $Q_{1H}^S= \frac{1-(\overline{x}^S)^2}{2}$. In anticipation of $Q^C$ and $Q_{1H}$, creators will form an expectation that the expected trining income they will receive is $\frac{Q^C}{Q_{1H}}f$, resulting in utility function of human creation:
\begin{align}
u_H^P(x) = \left(\beta + \frac{Q^C}{Q_{1H}^S}f\right) x - c_H.
\end{align}
By considering $u_H^P(x)$ jointly with $u_A(x)$ and $u_N(x) = 0$, we can determine the specific form of $(\underline{x}^S, \overline{x}^S)$ accordingly.  To prepare for determining these thresholds, we first show that $x_{HO}^P$ defined in Equation~\eqref{eq:x_HO_P} uniquely exists. To show that, define function $g_{HO}^P(x)$ based on the left hand side of Equation~\eqref{eq:x_HO_P}, that is,
\begin{align}\label{eq:g_HO_P}
g_{HO}^P(x) = \left[\beta + \frac{2 Q^C}{2 Q_0 + (1-x^2)}f \right] x - c_H, 
\end{align}
We can verify that $g_{HO}^P(x)$ increases in $x$. Further, note that 
\begin{align}
g_{HO}^P(x_{HO}^F) =  \left[\beta + \frac{2Q^C}{2Q_0 + 1-\left(x_{HO}^F\right)^2}f \right] x_{HO}^F - c_H
\end{align}
By the above definition of $Q_{1H}^F$, we have:  $Q_{1H}^F < \frac{1-\left(x_{HO}^F\right)^2}{2}$. Further, as $X_C < X_F$, we have $Q^C < Q_0 + Q_{1H}^F$. With $x_{HO}^F = \frac{c_H}{\beta+f}$, we have:
\begin{align}
g_{HO}^P(x_{HO}^F) <  \left[\beta + \frac{Q^C}{Q_0 + Q_{1H}^F}f \right] \left(\frac{c_H}{\beta+f} \right) - c_H < 0.
\end{align}
On the other hand, since $\beta > c_H$, we have:
\begin{align}
g_{HO}^P(1) = \beta + \frac{Q^C}{Q_0}f - c_H > 0.
\end{align}
Combined, we conclude that there exists a unique $x$, where $x \in (x_{HO}^F, 1)$, that satisfies $g_{HO}^F(x) = 0$, which is determined by Equation~\eqref{eq:x_HO_P}.

Similarly, we can show that $x_{HA}^P$ defined in Equation~\eqref{eq:x_HA_P} uniquely exists. Specifically, define function 
\begin{align}\label{eq:g_HA_P}
g_{HA}^P(x) = \left[\beta - (1-\lambda)\phi \beta + \frac{2Q^C}{2 Q_0 + (1-x^2)} f \right] x - \lambda \phi \beta X_1 - (c_H - c_A).
\end{align}
It can be verified that $g_{HA}^P(x)$ increases in $x$ for $x \in [0, 1]$. Further, we can show that $g_{HA}^P(1) \ge 0$ if and only if $X_1 < X_H^P$, where 
\begin{align}
X_H^P := \frac{\left[\beta - (1-\lambda)\phi \beta + \frac{Q^C}{Q_0} f \right] - (c_H - c_A)}{\lambda \phi \beta}. 
\end{align} 
On the other hand, by the definition of $x_{HA}^F$, we can also verify that for $X_1 < X_H^P$, $x_{HA}^F < 1$, and 
\begin{align}
g_{HA}^P(x_{HA}^F) \le \left[\beta - (1-\lambda)\phi \beta + \frac{Q^C}{Q_0 + Q_{1H}^F} f \right] \frac{c_H - c_A + \lambda \phi \beta X_1}{[1 - (1-\lambda)\phi]\beta + f} - \lambda \phi \beta X_1 - (c_H - c_A) < 0,
\end{align}
where the first inequality holds because $1-(x_{HA}^F)^2 \le Q_{1H}^F$, and the second inequality holds because $X^C < X^F$, and thus $Q^C < Q_0 + Q_{1H}^F$. Combined, we can show that when $X_1 < X_H^P$, there exists a unique $x \in (x_{HA}^F, 1)$ such that $g_{HA}^P = 0$, and this solution is defined in Equation~\eqref{eq:x_HA_P}.

Next, we establish the relative magnitude of $x_{HA}^P$ and $x_{HO}^P$. To see that, first note that when $X \ge X_H^P$, $x_{HA}^P = 1 > x_{HO}^P$. For $X < X_H^P$, by considering $g_{HA}^P(x)$, we have that $x_{HA}^P$ increases in $X_1$. On the other hand, $x_{HO}^P$ is independent of $X_1$. Further, we can verify that at $X_1 = X_A^P$, $g_{HA}^P(x_{HO}^P) = 0$. Therefore, $x_{HA}^P \le x_{HO}^P$ if and only if $X_1 \le X_A^P$.  

With the above analysis, we characterize thresholds $\underline{x}^S$ and $\overline{x}^S$ as follows.
\begin{enumerate}
\item When $X_1 \le X_A^P$, as discussed above, we have $x_{HA}^P \le x_{HO}^P$. Further, we can verify that with the expectation that the amount of human content created in Period-1 is $Q_{1H}^S = \frac{1-(x_{HO}^P)^2}{2}$, for creators with $x \ge x_{HO}^P$, $u_H^P(x) \ge \max(u_A(x), 0)$. Therefore, they choose human creation. On the other hand, for creators with $x < x_{HO}^P$, $0 \ge \max(u_H(x), u_A(x))$. Thus, they choose not to produce any content. Clearly, under this policy, the total amount of human data created in Period-1 is indeed $Q_{1H}^S$, consistent with the creators' expectation. Therefore, the creator policy with thresholds $\underline{x}^S = \overline{x}^S = x_{HO}^P$ is the equilibrium one. 
\item When $X_1 > X_A^P$, we have $x_{HA}^P > x_{HO}^P$. Further, we can verify that with the expectation that the amount of human content created in Period-1 is $Q_{1H}^S = \frac{1-(x_{HA}^P)^2}{2}$, for creators with $x \ge x_{HA}^P$, $u_H^P(x) \ge \max(u_A(x), 0)$. Therefore, they choose human creation. Under this policy, the total amount of human data created in Period-1 is indeed $Q_{1H}^S$, consistent with the creators' expectation. Therefore, the human creation policy with $\overline{x} = x_{HA}^P$ is the equilibrium policy. Accordingly, for creators with $x \le \max(x_{AO}, 0)$, we can verify that $0 \ge \max(u_A(x), u_H^P(x))$. Thus, $\underline{x}^S = \max(x_{AO}, 0)$. 
\end{enumerate}
This corresponds the Case 2 in the Appendix. \hfill $\square$

\medskip

\subsubsection*{Proof of Proposition \ref{prop:abundant_fair_use}.} First, by Proposition \ref{prop:model_improvement_generous}, we have that when $f = 0$ and $Q_0 \to +\infty$, the improved model quality $X_2 = 1$ . On the other hand, as $\lim_{X_2 \to 1} Q(X_2, X_1) = \infty$ (by Assumption \ref{ass:q}), for $f > 0$, we have $X_2 < 1$. Therefore, generous fair use ($f = 0$) maximizes model quality ($X_2$). 

Next, note that by Corollary \ref{coro:second_period_performance}, under a fixed $p_2$, the amount of (high-quality) AI content and creator income ($u_2$) both increase in $X_2$. Further, given $X_2$, both the amount of (high-quality) AI content and creator income ($u_2$) decreases in $p_2$. By Proposition \ref{prop:model_improvement_generous}, under generous fair use ($f = 0$), $p_2 = 0$. Therefore, both the amount of (high-quality) AI content and creator income ($u_2$) are both maximized under generous fair use.

Finally, we consider the Period-2 social welfare ($w_2$). Based on Equation~\eqref{eq:w_2} and the proof in Corollary \ref{coro:second_period_performance}, we re-write $w_2$ as a function of $(X_2, p_2)$:
\begin{align}
w_2(X_2, p_2) = \int_{\underline{x}(X_2, p_2)}^{\overline{x}(X_2, p_2)} [\lambda X_2 + (1-\lambda) x - c_A] + \int_{\overline{x}(X_2, p_2)}^{1} (x - c_H) d x;
\end{align}
where $\overline{x}()$ and $\underline{x}()$ are defined in Table \ref{table:second_period_threshold}. By definition, we can verify that $w_2(X_2, p_2)$ is continuous and differentiable on $X_2$ and $p_2$. 

Further, note that when $X_A(0) \ge 1$, that is, $c_A \ge \phi [\lambda \beta + (1-\lambda)c_H$, AI creation will not be adopted at any $(X_2, p_2)$. Thus, generous fair use weakly dominate all other fair use standard. In this proof, We focus on the interesting case with $X_A(0) < 1$, that is  
\begin{align}\label{eq:X_A_less_1}
c_A < \phi [\lambda \beta + (1-\lambda)c_H.    
\end{align}
Further, according to Table \ref{table:second_period_threshold}, the exact form of $\overline{x}$ and $underline{x}$ depends the magnitude of $\phi$. In this proof, we focus on the case of $\phi > \frac{\beta - c_H}{\beta(1-\lambda)}$. The proof with $\phi \le \frac{\beta - c_H}{\beta(1-\lambda)}$ follows similarly. We prove the result with four steps. 

\noindent {\bf Claim 1: $w_2(1, p_2) \ge w_2(X_2, p_2)$ for any $p_2$ and $X_2 \ge \min(1, X_H(p_2))$.}  

To show this claim, note that when $X_2 > \min(1, X_H(p_2))$, $\overline{x} = 1$, and thus
\begin{align}
w_2(X_2, p_2) = \int_{\underline{x}(X_2, p_2)}^{1} [\lambda X_2 + (1-\lambda) x - c_A].
\end{align}
Accordingly,
\begin{align}
\frac{\partial}{\partial X_2} w_2 = \lambda (1 - \underline{x}(X_2, p_2)) - [\lambda X_2 + (1-\lambda) \underline{x}(X_2, p_2) - c_A] \frac{\partial \underline{x}}{\partial X_2}.
\end{align}
By definition, we have $\underline{x}(X_2,p_2) \le 1$, so the first term is positive. For the second term, if $X_2 > X_N(p_2)$, then $\frac{\partial \underline{x}}{\partial X_2} = 0$. Otherwise, $\lambda X_2 + (1-\lambda) \underline{x}(X_2, p_2) - c_A > 0$ and $\frac{\partial \underline{x}}{\partial X_2} = -\frac{\lambda}{1-\lambda} < 0$. In each case, the second term is non-negative. Combined, we have $\frac{\partial}{\partial X_2} w_2 \ge 0$, as desired.

\noindent {\bf Claim 2: $w_2(1, 0) > w_2(1, p_2)$ for $p_2 > 0$.} 

To prove this claim, consider two scenarios depending on the magnitude of $X_H(p)$.
\begin{enumerate}
\item if $X_H(p) > 1$, $\overline{x} = x_{HA} < 1$, then
\begin{align}
w_2(1, p_2) = \int_{x_{AO}(1, p_2)}^{x_{HA}(1, p_2)} [\lambda + (1-\lambda) x - c_A] + \int_{x_{HA}(1, p_2)}^{1} (x - c_H) d x;
\end{align}
Take derivative with respect to $p_2$, we have that $w_2(1, p_2)$ decreases in $p_2$. 
\item if $X_H(p) \le 1$, $\overline{x} = 1$, and $w_2(1, p_2)$ also decreases in $p_2$. 
\end{enumerate}
Combined, we have that $w_2(1,0) > w_2(1, p_2)$ for any $p_2 > 0$. 

\noindent {\bf Claim 3: $w_2$ is convex in $X_2$ when $X_2 \in [\max(0, X_A(p_2)), \min(1, X_H(p_2))]$.}

To show this claim, note that when $X_2 \in [\max(0, X_A(p_2)), \min(1, X_H(p_2))]$, $\overline{x}(X_2, p_2) = x_{HA}(X_2, p_2)$ and $\underline{x}(X_2, p_2) = x_{AO}(X_2, p_2)$. For notational convenience, we suppress the dependence of $x_{HA}$ and $x_{AO}$ on $X_2$ and $p_2$. By Table \ref{table:second_period_threshold}, we can verify that $0 < x_{AO} < x_{HA} < 1$. By substituting $x_{HA}$ and $x_{AO}$ into $w_2(X_2, p_2)$, we have:
\begin{align}
w_2(X_2, p_2) = \int_{x_{AO}(X_2, p_2)}^{x_{HA}(X_2, p_2)} [\lambda X_2 + (1-\lambda) x - c_A] + \int_{x_{HA}(X_2, p_2)}^{1} (x - c_H) d x.
\end{align}
Take partial derivative of $w_2$ with respect to $X_2$ within this range, we have:
\begin{align}
\frac{\partial}{\partial X_2} w_2 &= \frac{\partial}{\partial X_2} \int_{x_{AO}(X_2, p_2)}^{x_{HA}(X_2, p_2)} [\lambda X_2 + (1-\lambda) x - c_A] d x + \frac{\partial}{\partial X_2} \int_{x_{HA}(X_2, p_2)}^{1} (x - c_H) d x \\
& = \lambda (x_{HA} - x_{AO}) + [\lambda (X_2 - x_{HA}) + c_H - c_A] \frac{\partial x_{HA}}{\partial X_2} - [\lambda X_2 + (1-\lambda)x_{AO} - c_A] \frac{\partial x_{AO}}{\partial X_2}
\end{align}
As $\frac{\partial x_{AO}}{\partial X_2} = -\frac{\lambda}{1-\lambda} < 0$ and $\frac{\partial x_{HA}}{\partial X_2} = \frac{\lambda \phi}{1-(1-\lambda)\phi} \in (0, 1)$, we have:
\begin{align}
\frac{\partial^2}{\partial X_2^2} w_2 & = \lambda \left(\frac{\partial x_{HA}}{\partial X_2} - \frac{\partial x_{AO}}{\partial X_2}\right) + \left[\lambda\left(1-\frac{\partial x_{HA}}{\partial X_2}\right)\right] \frac{\partial x_{HA}}{\partial X_2} - \left[\lambda + (1-\lambda)\frac{\partial x_{AO}}{\partial X_2}\right]\frac{\partial x_{AO}}{\partial X_2}.
\end{align}
Substituting $\frac{\partial x_{AO}}{\partial X_2} = -\frac{\lambda}{1-\lambda}$ and $\frac{\partial x_{HA}}{\partial X_2} = \frac{\lambda \phi}{1-(1-\lambda)\phi}$ into the above equation, we have $\frac{\partial^2}{\partial X_2^2} w_2 > 0$. 

\noindent {\bf Claim 4: $w_2(1, 0) \ge w_2(X_2, p_2)$ for any $X_2 < 1$ or $p_2 > 0$.}

To prove this claim, according to Claim 3, by convexity, for any $X_2 \in [\max(0, X_A(p_2)), \min(1, X_H(p_2))]$, $w_2$ takes its maximum at either $X_2 = \max(0, X_A(p_2))$ or $X_2 = \min(1, X_H(p_2))$. Further, we have shown in Claim 1 that $w_2(1, p_2) > w_2(X_2, p_2)$ for any $X_2 \ge \min(1, X_H(p_2))$. Combining this with Claim 2, we have $w_2(1, 0) > w_2(\min(1, X_H(p_2)),p_2)$ for any $p_2$. Therefore, to prove Claim 4, we only need to prove that $w_2(1,0) > w(\max(0, X_A(p_2)), p_2)$. To see that, we first calculate $w_2(1,0)$, which follows:
\begin{align}
w_2(1, 0) 
&= \int_{x_{AO}(1,0)}^{x_{HA}(1,0)} [\lambda + (1-\lambda) x - c_A] d x + \int_{x_{HA}(1,0)}^{1} (x - c_H) dx
\end{align}
To compare this with $w_2(\max(0, X_A(p_2)), p_2)$, we consider two scenarios.
\begin{enumerate}
\item When $X_A(p_2) \ge 0$, we have $\underline{x} = \overline{x} = x_{HO}$, and therefore,
\begin{align}
w_2(X_A, p_2) &= \int_{x_{HO}}^{1} (x - c_H) d x; 
\end{align}
Comparing that with $w_2(1,0)$, we have:
\begin{align}
w_2(1,0) - w_2(X_A, p_2) = \int_{x_{AO}(1, 0)}^{x_{HO}} [\lambda + (1-\lambda) x - c_A] d x + \int_{x_{HO}}^{x_{HA}(1,0)} [\lambda (1 - x) + c_H - c_A] d x
\end{align}
It is clear that both integrands are positive. Further, by Eq.~\eqref{eq:X_A_less_1}, we can verify that $x_{AO}(1,0) < x_{HO} < x_{HA}(1,0)$. Thus, $w_2(1,0) > w_2(X_A, p_2)$.
\item When $X_A(p_2) \le 0$, we compare $w_2(1,0)$ with $w_2(0, p_2)$. To calculate $w_2(0, p_2)$, we note that $x_{HA}(0, p_2) = \frac{c_H - c_A - p}{[1-(1-\lambda)\phi]\beta}$, and $x_{AO}(0, p_2) = \frac{c_A - p}{(1-\lambda)\phi \beta}$. Therefore, 
\begin{align}
w_2(0, p_2) = \int_{x_{AO}(0, p_2)}^{x_{HA}(0, p_2)} [(1-\lambda) x - c_A] + \int_{x_{HA}(0, p_2)}^{1} (x - c_H) d x.
\end{align}
\end{enumerate}
It is easy to verify that $x_{AO}(1,0) < x_{AO}(0,p_2) < x_{HA}(0, p_2) < x_{HA}(1,0)$. Therefore,
\begin{align}
w_2(1, 0) - w_2(0, p_2) = \int_{x_{AO}(1,0)}^{x_{AO}(0,p_2)} [\lambda + (1-\lambda) x - c_A] d x + \int_{x_{AO}(0,p_2)}^{x_{HA}(0,p_2)} \lambda  d x +\int_{x_{HA}(0,p_2)}^{x_{HA}(1,0)} [\lambda(1-x) +c_H - c_A] dx.
\end{align}
It is easy to verify that the second and third terms are both positive. For the first term, we can see that for $x > x_{AO}(1,0)$, the integrand is also positive. Thus $w_2(1, 0) > w_2(0, p_2)$.

Combining the above four steps, we have shown that $w_2(1,0) > w_2(X_2, p_2)$ for any $X_2 < 1$ or $p_2 \ge 0$. By Proposition \ref{prop:model_improvement_generous}, we have $X_2 = 1$ and $p_2 = 0$ at $f = 0$.  \hfill $\square$

\medskip

\subsubsection*{Proof of Proposition \ref{prop:abundant_copyrightability_1}.} Note from the expressions in Proposition \ref{prop:second_period_production},  we have that $\overline{x}$ increases in $\phi$ and $\underline{x}$ decreases in $\phi$. Next, by the expressions in Corollary \ref{coro:second_period_performance}, $Q_{2A}$ and $r_2$ increase in $\phi$, and $Q_{2H}$ decreases in $\phi$.  Regarding $u_2$, by writing it as 
\begin{align}
u_2 = \int_0^1 \max(u_H, u_A, 0) dx
\end{align}
we observe that as $u_A$ increases in $\phi$ and $u_H$ is independent of $\phi$, we have $u_2$ increases in $\phi$.  

For social welfare, we note that
\begin{align}
w_1 = \int_{\underline{x}}^{\overline{x}} [\lambda X_1 + (1-\lambda) x - c_A] + \int_{\overline{x}}^{1} (x - c_H) d x;
\end{align}
take derivative of social welfare with respect to $\phi$.
\begin{align}
\frac{\partial w_2}{\partial \phi} &= \frac{\partial \overline{x}}{\partial \phi} \left[\lambda X_1 + (1-\lambda)\overline{x} - c_A\right] -  \frac{\partial \underline{x}}{\partial \phi} \left[\lambda X_1 + (1-\lambda)\underline{x} - c_A\right] - \frac{\partial \overline{x}}{\partial \phi} \left(\overline{x} - c_H\right) \\
& = \frac{\partial \overline{x}}{\partial \phi} \left[\lambda (X_1 - \overline{x}) +c_H - c_A\right] - \frac{\partial \underline{x}}{\partial \phi} \left[\lambda X_1 + (1-\lambda)\underline{x} - c_A\right]
\end{align}
Note that $\frac{\partial \underline{x}}{\partial \phi} \le 0$, and $\lambda X_1 + (1-\lambda)\underline{x} - c_A > 0$. Therefore, the second term is positive. For the first term, note that when $X_1 > X_H(0)$, $\frac{\partial \overline{x}}{\partial \phi} = 0$. Therefore $\frac{\partial w_2}{\partial \phi} > 0$, as desired. \hfill $\square$

\medskip

\subsubsection*{Proof of Proposition \ref{prop:abundant_copyrightability_2}.}
Based on the proof of Corollary \ref{coro:second_period_performance}, to prove that the amount of Period-2 AI content and Period-2 creator income increase in $\phi$, it is sufficient to show that the creators' utility of AI creation $u_A = \phi \beta [\lambda X_2 + (1-\lambda)x] -c_A - p_2$ increase in $\phi$. To prove this, consider two scenarios: (i) generous fair use ($f =0$); and (ii) strict fair use ($f > 0$). 

Under generous fair use ($f = 0$), according to Proposition \ref{prop:model_improvement_generous}, $X_2 = 1$ and $p_2 = 0$. Therefore, $u_A$ increases in $\phi$, as desired. 

Under strict fair use ($f > 0$), a sufficient condition $u_A$ to increase in $\phi$ is that $v_A := \phi \beta \lambda X_2 - p_2$ increases in $\phi$. To see this, note that $v_A$ is the objective function that the AI company maximizes, where $p_2 = p_2^X(X_2)$ Note that this objective function is not only directly affected by $\phi$, but also indirectly through $X_2$ and $p_2$. We prove that this function increases in $\phi$ in two steps.

First, we show that $p_2^X(X)$ is non-increasing in $\phi$. To see that, for a fixed $X$, define the feasible price set of $p_2$ as follows:
\begin{align}
S_X(\phi) := \{p: r_2(X, p; \phi) - f q(X) \ge 0.\}
\end{align}
By Corollary \ref{coro:second_period_performance}, we can show that $r_2$ increases in $\phi$. Therefore, for any $\phi_2 > \phi_1$, we have $S_X(\phi_1) \subseteq S_X(\phi_2)$. As $p_2^X(X; \phi) = \min S_X(\phi)$ by definition, we have that for $\phi_2 > \phi_1$, $p_2^X(X; \phi_2) < p_2^X(X; \phi_1)$.

Next, to show that $v_A(X; \phi)$ increases in $\phi$. Take any $\phi_2 > \phi_1$, we have that $\Omega_2(\phi_1) \subseteq \Omega_2(\phi_2)$. Let $X_1 \in \Omega_2(\phi_1)$ be the optimal $X$ that maximizes $v_A(X; \phi_1)$, then we have:
\begin{align}
v_A(X_1; \phi_1) = \phi_1 \beta \lambda X_1 - p_2^X(X_1; \phi_1) < \phi_2 \beta \lambda X_1 - p_2^X(X_1; \phi_2) = v_A(X_1; \phi_2).
\end{align}
Further, let $X_2 \in \Omega_2(\phi_2)$ be the optimal $X$ that maximizes $v_A(X; \phi_2)$. By definition, we have $v_A(X_2; \phi_2) \ge v_A(X_1; \phi_2)$. Therefore, we have $v_A(X_1; \phi_1) < v_A(X_2; \phi_2)$, as desired. \hfill $\square$

\medskip

\subsubsection*{Proof of Corollary \ref{coro:abundant_interaction}.} According to Proposition \ref{prop:abundant_fair_use}, regardless of AI-copyrightability ($\phi$), generous fair use ($f=0$) maximizes social welfare and creator income in Period-2. As fair use standard has no impact on Period-1 performance metrics, combining the two periods, generous fair use maximizes long-term social welfare and creator income. 

For AI-copyrightability, by Proposition \ref{prop:abundant_copyrightability_1}, when $X_1$ is sufficiently large, both Period-1 social welfare and creator income increase in $\phi$. Thus, full AI-copyrightability is optimal. For Period-2, under generous fair use, $X_2 = 1$ and $p_2 = 0$. Similar to the proof of Proposition \ref{prop:abundant_copyrightability_1}, we can show that full AI-copyrightability maximizes Period-2 social welfare and creator income. Combined, full AI-copyrightability maximizes long-term social welfare and creator incomes.  \hfill $\square$

\medskip

\subsubsection*{Proof of Proposition \ref{prop:scarce_fair_use_generous}.} When $X_1 > X_H$, no human content is produced in Period-1 under generous fair use. Thus, the improved model quality $X_2 = X_1$. On the other hand, under strict fair use, the improved model quality $X_2 \ge X_1$, as desired. \hfill $\square$

\medskip

\subsubsection*{Proof of Proposition \ref{prop:scarce_fair_use_strict}.} When $f$ is sufficiently large, by considering Proposition \ref{prop:model_improvement_strict_large_X} and \ref{prop:abundant_copyrightability_2}, $\pi_2(X_2, p_2) < 0$ for all $X_2 > X_1$. Thus, the AI company chooses not to improve the model. As a result, the amount of data required for model training $Q^C = 0$. By Appendix \ref{appx:first_period_thresholds}, $\overline{x}^S = \overline{x}(X_1, 0)$ and $\underline{x}^S = \underline{x}(X_1, 0)$, the same as the policy under generous fair use (Proposition \ref{prop:first_period_abundant}).

For Period-2, under strict fair use (high $f$), the improved model quality, let it be $X_2^S$, equals to $X_1$. In contrast, under generous fair use ($f = 0$), by Proposition \ref{prop:model_improvement_generous}, the AI company utilize all human content created in Period-1 ($\frac{1-(\overline{x}^S)^2}{2}$), and thus the improved model quality is $X_2^G \ge X_1 = X_2^S$. Further, in both cases, $p_2 = 0$. For creator incomes, following the proof of Corollary \ref{coro:second_period_performance}, we can show that creator incomes are higher under $(X_2^G, p_2 = 0)$ than that under $(X_2^S, 0)$. 

For social welfare, we have:
\begin{align}
w_2(X) &= \int_{\underline{x}}^{\overline{x}} [\lambda X_2 + (1-\lambda) x - c_A] dx + \int_{\overline{x}}^{1} (x- c_H) dx;
\end{align}
and thus,
\begin{align}
\frac{\partial}{\partial X_2}w_2 = \lambda (\overline{x} - \underline{x}) - \frac{\partial \underline{x}}{\partial X} [\lambda X + (1-\lambda) \underline{x} - c_A] + \frac{\partial \overline{x}}{\partial X} [\lambda (X - \overline{x}) + c_H - c_A].
\end{align}
Note that $\overline{x} \ge \underline{x}$, $\underline{x}$ is non-increasing in $X_2$. Further, by the definition of $\underline{x}$, we have $\lambda X + (1-\lambda) \underline{x} - c_A > 0$, thus, a sufficient condition for $\frac{\partial w_2}{\partial X} \ge 0$ is 
\begin{align}\label{eq:overline_x_X}
\frac{\partial \overline{x}}{\partial X} [\lambda (X - \overline{x}) + c_H - c_A] \ge 0.
\end{align}
By Table \ref{table:second_period_threshold}, when $X \ge X_H(0)$, $\frac{\partial \overline{x}}{\partial X} = 0$, which satisfies Equation~\eqref{eq:overline_x_X}. For $X < X_H$, $\frac{\partial \overline{x}}{\partial X} > 0$, Equation~\eqref{eq:overline_x_X} is satisfied if $\lambda (X_2 - x_{HA}) + c_H - c_A \ge 0.$ Or equivalently,
\begin{align}
X \ge X^{SW} := \left(1 - \frac{(1-\phi)\beta}{\lambda(1-\phi \beta)}\right) (c_H - c_A). 
\end{align}
Therefore, when $X_1 \ge \min(X_H(0), X^{SW})$, we have $w_2(X_2^G, 0) = w_2(X_1, 0) \le w_2(X_2^S, 0)$. \hfill $\square$

\medskip

\subsubsection*{Proof of Proposition \ref{prop:scarce_copyrightability}.} According to the proof of Corollary \ref{coro:second_period_performance}, the amount of AI content in Period-1 ($Q_{1A}$) and Period-1 creator income ($u_1$) both increase in $\phi$, but the amount of human content in Period-1 ($Q_{1H}$) decreases in $\phi$. Further, by Proposition \ref{prop:model_improvement_generous}, under generous fair use, the AI company uses up all of $Q_{1H}$ for model training. As a result, $X_2$ decreases in $\phi$. \hfill $\square$

\medskip

\end{APPENDICES}
\end{document}